\documentclass[aps, prd, amsmath, floats, floatfix, twocolumn, superscriptaddress, nofootinbib, showpacs]{revtex4-1}

\usepackage[utf8]{inputenc}
\usepackage[caption=false,labelformat=empty]{subfig}
\usepackage{graphicx}
\usepackage{xcolor}
\usepackage{xspace}
\usepackage{soul}
\usepackage{afterpage}
\usepackage{subcaption}
\usepackage{url}
\usepackage{bm}
\usepackage{times}
\usepackage{acronym}
\usepackage{dcolumn}
\usepackage{epsf}
\usepackage{amssymb}
\usepackage{float}
\usepackage{hyperref}
\usepackage{comment}
\usepackage{array}
\usepackage{natbib}
\normalsize
\usepackage{booktabs}
\usepackage{tabularx}
\allowdisplaybreaks
\usepackage{multirow}
\usepackage{afterpage}
\usepackage{aas_macros}

\newacro{GW}[GW]{gravitational wave}
\newacro{QC}[QC]{quasi-circular}
\newacro{BF}[BF]{Bayes factor}
\newacro{BBH}[BBH]{binary black hole}
\newacro{LVK}[LVK]{LIGO-Virgo-KAGRA Collaboration}
\newacro{BH}[BH]{black hole\acused{BHBH}\acused{NSBH}}
\newacro{NS}[NS]{neutron star\acused{NSNS}\acused{NSBH}}
\newacro{BHBH}[BH-BH]{\acl{BH}-\aclu{BH}}
\newacro{NSBH}[NS-BH]{\aclu{NS}-\aclu{BH}}
\newacro{NSNS}[NS-NS]{\acl{NS}-\aclu{NS}}
\newacro{EOS}[EOS]{equation of state}
\acrodefplural{EOS}{equations of state}
\newacro{GRB}[GRB]{gamma-ray burst}
\newacro{EM}[EM]{electromagnetic}
\newacro{PE}[PE]{parameter estimation}
\newacro{EOB}[EOB]{effective one body}
\newacro{NR}[NR]{numerical relativity}
\newacro{AdV}[AdV]{Advanced Virgo}
\newacro{aLIGO}[aLIGO]{Advanced LIGO}
\newacro{pn}[PN]{post-Newtonian}
\newacro{NR}[NR]{numerical relativity}
\newacro{PISn}[PISn]{pair-instability supernova}
\newacro{cbc}[CBC]{compact binary coalescence}
\newacro{nqc}[NQC]{next to quasi-circular}
\newacro{ce}[CE]{Cosmic Explorer}
\newacro{et}[ET]{Einstein Telescope}
\newacro{GWOSC}[GWOSC]{Gravitational Wave Open Science Center}
\newacro{gwtc}[GWTC]{Gravitational-Wave Transient Catalog}
\newacro{ci}[CI]{credible interval}
\newacro{snr}[SNR]{signal-to-noise ratio}

\newcommand{\bmath}[1]{{\boldmath #1 \unboldmath}}

\newcommand\qmstateproduct[2]{\left<#1|#2\right>}

\newcommand\dali[0]{{\texttt{TEOBResumS-Dalí}}}
\newcommand{\totalevents}{38 }

\newcommand{\RIFT}{\texttt{RIFT}\xspace}

\newcommand{\nrsur}{\texttt{NRSur7dq4}}
\newcommand{\chieff}{\chi_{\rm eff}}

\newcommand{\beq}{\begin{equation}}
\newcommand{\eeq}{\end{equation}}
\newcommand{\beqn}{\begin{eqnarray}}
\newcommand{\eeqn}{\end{eqnarray}}

\newcommand{\INFN}{\affiliation{INFN Sezione di Torino, Via P. Giuria 1, 10125 Torino, Italy}}
\newcommand{\UT}{\affiliation{Dipartimento di Fisica, Universit\`a di Torino, via P. Giuria 1, 10125 Torino, Italy }}
\newcommand{\RIT}{\affiliation{Center for Computational Relativity and Gravitation, Rochester Institute of Technology, Rochester, NY 14623, USA}}
\newcommand{\PU}{\affiliation{Phenikaa Institute for Advanced Study, Phenikaa University, Hanoi, Vietnam}}
\newcommand{\GT}{\affiliation{School of Physics and Center for Relativistic Astrophysics, Georgia Institute of Technology, Atlanta, GA 30332, USA}}

\newcommand{\Ken}{\affiliation{Department of Physics, Kennesaw State University, Marietta, GA 30060, USA}}
\begin{document}
\raggedbottom
\title{Gravitational Wave Hyperbolic Catalog: Reanalyzing High-Mass Gravitational Wave Signals Using Hyperbolic Waveforms}
\author{Jacob Lange}
\INFN
\author{Danilo Chiaramello}
\UT
\INFN
\author{Peter Lott}
\PU
\author{Chad Henshaw}
\Ken
\GT
\author{Alessandro Nagar}
\INFN
\author{Richard O'Shaughnessy}
\RIT
\author{Laura Cadonati}
\GT
\begin{abstract}

Close hyperbolic encounters between black holes produce distinctive bursts of gravitational radiation with a time-frequency morphology that is qualitatively different from that of quasi-circular inspirals.
Expected to arise in dense stellar environments through dynamical interactions, these encounters probe formation channels and mass ranges inaccessible to isolated binary evolution, making them a compelling target for current and next-generation detectors.
In this work, we reanalyze \totalevents high-mass events from the LIGO-Virgo-KAGRA catalogs using the hyperbolic configuration of the~\dali~waveform model.
We compare these with analyses using the quasi-circular, precessing configuration of the same model, computing Bayes factors to evaluate which description is favored by the data.
We find that most events strongly to mildly favor the quasi-circular, precessing scenario, except for GW190521.
For this event, we find that the signal is best fit by a dynamical capture waveform, with Bayes factor $\ln \mathcal{B}^{\rm hyp}_{\rm prec}=3.71^{+0.11}_{-0.11}$.
We confirm this preference via further analyses with~\dali~in different configurations (quasi-circular, non-precessing; eccentric, non-precessing; and eccentric, precessing), as well as one using the quasi-circular, precessing numerical relativity surrogate model \nrsur.
We also highlight the results we obtain for GW231123, another high-mass signal linked to evidence of strong precession,
for which we find strong preference for the quasi-circular, precessing scenario, with $\ln \mathcal{B}^{\rm hyp}_{\rm prec}=-15.80^{+0.24}_{-0.24}$.
To further build confidence in our results, we also analyze synthetic signals built from the maximum likelihood points in our hyperbolic and quasi-circular, precessing analyses for events GW190521 and GW231123.
While the two GW231123-like signals strongly favored analyses in their corresponding scenario, the GW190521-like signals did so only marginally.
\end{abstract}

\maketitle

\section{Introduction}
\label{sec:intro}
Over 200 \ac{GW} signals have been reported so far by the \ac{LVK}~\cite{LIGOScientific:2025slb}.
The majority of these astrophysical events are associated with \ac{BBH} mergers in bound systems; several binaries involving neutron stars have also been detected~\cite{LIGOScientific:2018cki, LIGOScientific:2021qlt}.
Pulsar timing observations have led to the detection of a stochastic gravitational-wave (GW) background, widely interpreted as originating from an ensemble of supermassive \ac{BH} binaries \cite{NANOGrav:2023gor}.
\ac{GW} observations by current observatories provide insight into the strong-field regime of gravity, allowing  tests of general relativity\cite{LIGOScientific:2020tif, LIGOScientific:2021sio}, observation of the population mass profile~\cite{LIGOScientific:2020tif, LIGOScientific:2021sio}, and constraints on the neutron star equation of state~\cite{LIGOScientific:2018cki}, among other pursuits~\cite{LIGOScientific:2018mvr, Wette:2023lqi, LIGOScientific:2025rid}.

Binary bound systems that culminate in merger, collectively known as \acp{cbc}, are ideal sources for ground-based laser interferometers~\cite{Finn:1992xs} because the characteristic frequencies of lighter binaries typically fall within their 10–1000 Hz sensitivity band.
However, current observatories have limited sensitivity at the lower end of this range, creating ambiguity in detections unless systematics are properly modeled~\cite{Capote:2024rmo}.
This low-frequency regime is nonetheless important, as it opens the possibility of observing heavier-mass binaries, such as GW190521~\cite{LIGOScientific:2020iuh} and GW231123~\cite{LIGOScientific:2025rsn}.
In particular, in the sub-hertz ($<1$ Hz) band, one may even detect binaries on highly eccentric ($e > 0.8$) or fly-by orbits, including those involving \acp{BH}.
These events may result in several outcomes: they may be hyperbolic encounters, meaning they scatter once without subsequent interaction, or become bound in a highly eccentric orbit, resulting in multiple fly-bys and subsequent merger (\textit{zoom-whirl} orbit), or they may collapse directly into a \textit{plunge}.
These systems release gravitational radiation during close encounters as \textit{GW bremsstrahlung}~\cite{Peters1963, Hansen1972, 1977ApJ...216..610T,1978ApJ...224...62K}.
These dynamical systems have a rich morphological structure~\cite{Levin:2008ci, 2017PhRvD..96h4009B, 2024PhRvL.132z1401B} which becomes even richer with the inclusion of higher-order modes.
Moreover, the time-frequency content of \ac{GW} signals in this scenario is morphologically distinct compared to those from \ac{QC} systems~\cite{Morras:2021atg, Bini:2023gaj, Henshaw_2025, Lott:2025ydm}, manifesting in laser interferometers as characteristic blip-like profiles in time, with broadband frequency.

Although no detections have been confirmed to date, eccentric binaries may be relevant sources in densely packed clusters or near active galactic nuclei; it has been estimated that at least BBH hyperbolic encounters could be common given simulations and event rate estimates~\cite{Kocsis:2006hq,Capozziello:2008mn,Mukherjee:2020hnm}.
If detected, BBHs in hyperbolic or zoom-whirl orbits could therefore provide key insight into the population of \acp{BH} and other compact objects, and help detail a more complete picture of the stochastic gravitational-wave background~\cite{Garcia-Bellido:2021jlq, Kerachian:2023gsa}.
GW bremsstrahlung, detection rates peaking in the~1Hz range, is therefore relevant for A\#~\cite{KAGRA:2013rdx} and next generation detectors like \ac{ce}~\cite{Reitze:2019iox} and \ac{et}~\cite{Maggiore:2019uih}.

Dense stellar environments naturally assemble binaries, trigger close encounters, and produce hierarchical mergers that populate \ac{BH} masses inaccessible to single-star evolution \cite{OLeary2009,Samsing:2017xmd,Rodriguez:2017pec,Gerosa:2021mno}, accumulating \ac{BH} mergers within the~\textit{pair-instability mass gap}, whose boundaries are set by the physical stability of very massive stars ($\sim50 - 120 M_{\odot}$).
In this range, the heat inside the core can lead to photon conversion into electron-postitron pairs, reducing radiation pressure within the star.
This can in turn cause the violent ejection of much of the mass through oxygen burning, sometimes leading to complete disruption through pair-instability supernovae~\cite{Heger:2001cd,Woosley:2016hmi}.
Below this range of masses, stars are too small to trigger pair production and collapse normally, while above it, the cores are so massive that pair production only speeds up their direct collapse into \acp{BH}.
Dynamical formation channels are natural factories of binaries that may populate the otherwise forbidden pair-instability mass gap, making \ac{GW} bremsstrahlung a potentially important probe of mass-gap mergers~\cite{OLeary2009,Samsing:2017xmd,Rodriguez:2017pec,Gerosa:2021mno} as well as a source of information on the \ac{BBH} formation channels in dense environments~\cite{Li:2022qoz, 2025ApJ...990..210L}.
\cite{Garcia-Bellido:2017qal,Garcia-Bellido:2017knh} present the first computation of the power emitted from close hyperbolic encounters to first order.

\Ac{PE} techniques for gravitational-wave sources have almost exclusively focused on \ac{BBH} mergers in bound orbits, with a few exceptions.
The first analysis of a \ac{GW} signal to consider unbound orbits was presented in~\cite{Gamba:2021gap}, who used the \dali~waveform model in its hyperbolic configuration.
Assuming non-spinning \acp{BH} and using only the dominant quadrupolar modes $(2, \pm 2)$, the authors there analyzed the high-mass GW190521 event, finding the hyperbolic description to be favored by the data over a \ac{QC}, precessing one by a \ac{BF} of 4300 to 1, without taking astrophysical prior odds into account.
Additionally, \citet{Fontbute:2024amb} produced an accurate \ac{NR} surrogate for hyperbolic encounters of equal mass spin-aligned \acp{BBH} with higher order modes.
However, they struggled to correctly recover the parameters of synthetic signals in \ac{PE}.
More recently, some of the present authors showed in~\cite{Henshaw_2025} accurate recovery of the parameters of synthetic signals from direct plunges and scattering encounters analyzed with~\RIFT~\cite{Lange2018,PhysRevD.99.084026,PhysRevD.107.024040,Wagner:2025bih}, an algorithm that performs Rapid parameter inference on gravitational wave sources via Iterative FiTting, and modeled via~\dali~\cite{Chiaramello:2020ehz,Nagar:2020xsk,Nagar:2021gss,Nagar:2021xnh,Nagar:2024dzj,Nagar:2024oyk,Gamba:2024cvy,Albanesi:2025txj}.

Motivated by these two previous studies on hyperbolic \ac{PE}, we present in this work the re-analyses of~\totalevents~\ac{GW} signals sourced by \acp{BH} in the pair-instability mass gap using the hyperbolic and precessing, \ac{QC} configurations of~\dali.
For the each event, we include non-precessing spins in the hyperbolic analysis, as well as higher-order modes (specifically, the $(2,\pm2), (2,\pm1), (3,\pm3)$, and $(4,\pm4)$ modes in the co-precessing frame).
To strengthen our results, for two highlighted events, GW190521 and GW231123, we present additional analyses including eccentricity and/or spin-precession using~\dali, and precessing runs with the \texttt{NRSur7dq4}~\cite{2019PhRvR...1c3015V} model.

This paper is organized as follows.
In Section~\ref{sec:methods} we introduce the parameter space of \acp{BBH} on unbound orbits considered in this work, the \dali~\ac{GW} model, the \ac{PE} code used~\RIFT, and detail the settings we employ in the analyses.
In Section~\ref{sec:results}, we describe the results of our analyses, presenting our catalog of \totalevents events, including the deeper study of the two notable cases mentioned above.
We conclude in Section~\ref{sec:conclusions} with a summary and interpretations of our findings.

\paragraph*{Conventions}
Throughout this paper we use geometrized units with $G = c = 1$.
We consider \ac{BBH} systems with (redshifted, detector-frame) masses $m_1 > m_2$, with the total mass given by $M = m_1 + m_2$ and the mass ratio defined by $q = m_1/m_2 > 1$.
The source-frame total mass is found via $M_{\rm source} = (1+z)^{-1} M$, where $z$ is the redshift; similar relations hold for the components.
The symmetric mass ratio is $\nu = \mu/M = m_1 m_2/M^2$, where $\mu$ is the reduced mass.
The \acp{BH}' spin vectors are $\bm{S}_{1,2} = m_{1,2}^2 \bm{\chi}_{1,2}$, from which we define the effective aligned $\chi_{\rm eff}$ and precessing spin $\chi_{\rm  p}$ parameters~\cite{Racine:2008qv, Ajith:2009bn, PhysRevD.91.024043}:
\begin{subequations}
  \begin{align}
    \chi_{\rm eff} &= \dfrac{m_1 \bm{\chi}_1 + m_2 \bm{\chi}_2}{M} \cdot \hat{\bm{L}}\, , \label{eq:chieff}\\
    \chi_{\rm p}   &= \max \left\{
        |\bm{\chi}_{1}^\perp|, \dfrac{4+3q}{4q^2+3q} |\bm{\chi}_{2}^\perp| \label{eq:chip}
    \right\}\, ,
  \end{align}
\end{subequations}
where $\hat{\bm{L}} = \bm{L}/|\bm{L}|$ is the orbital angular momentum vector, and $\bm{\chi}_{1,2}^{\perp}$ are the spin vector components perpendicular to it.

\section{Methods}
\label{sec:methods}

\subsection{GW Models}
\label{subsec:models}
\dali~\cite{Chiaramello:2020ehz,Nagar:2020xsk,Nagar:2021gss,Nagar:2021xnh,Nagar:2024dzj,Nagar:2024oyk,Gamba:2024cvy,Albanesi:2025txj} is a state-of-the-art \ac{GW} model based on the \ac{EOB} formalism~\cite{Buonanno:1998gg,Damour:2001tu}.
It comprises three fundamental building blocks.
(i) The conservative two-body dynamics is mapped to the effective problem of a particle moving in an external spacetime, defined by a Hamiltonian $H_{\rm EOB}$, which resums perturbative \ac{pn} results and incorporates strong-field information from \ac{NR} simulations in order to achieve phasing accuracy down to the plunge and merger stages.
The relationship between the real and effective energy is one of the hallmarks of the \ac{EOB} approach:
\begin{equation}
    \label{eq:eob_ham}
    H_{\rm EOB} = M \sqrt{1 + 2 \nu \bigl(\hat{H}_{\rm eff} - 1\bigr)}\, .
\end{equation}
(ii) The emitted waveform is decomposed into multipoles $h_{\ell m}$, which are computed from the dynamics according to a factorized and resummed prescription~\cite{Damour:2008gu,Nagar:2019wds,Nagar:2020pcj}; the inspiral signal is completed with a phenomenological description of the merger-ringdown stage~\cite{Damour:2014yha}, stitched together around the time of the peak of the dominant (for \ac{QC} systems) $(2,2)$ mode.
A smooth match is facilitated by the \ac{nqc} corrections~\cite{Damour:2007xr,Damour:2009kr}, which improve the description of the plunge stage of the waveform by incorporating an explicit dependence on the radial momentum and acceleration, with coefficients determined by enforcing a $C^1$ match (in amplitude and frequency) with the \ac{NR}-informed merger-ringdown.
(iii) A radiation reaction force, built on the highly effective waveform model, parametrizes the dissipative effect of \ac{GW} emission onto the dynamics, driving the inspiral.
\dali~builds on the \texttt{TEOBResumS-GIOTTO}~model~\cite{Nagar:2018zoe,Akcay:2020qrj,Gamba:2021ydi}, which is specialized to the case of quasi-spherical inspirals, to cover \emph{generic} orbital configurations, including bound, eccentric systems as well as initially unbound ones, which can result in hyperbolic-like scattering encounters or dynamical captures eventually resulting in a merger.
This is naturally supported by the Hamiltonian formulation of the conservative dynamics, which is valid for any kind of (planar) orbit, and is made possible by Newtonian-level non-circular corrections introduced into the waveform and radiation reaction models during the inspiral, which do not rely on explicit assumptions about the orbital geometry.

Initial data for an unbound binary can be specified by the system's initial radial separation $r_0$, energy $E_0/M \equiv H_{\rm EOB} (r_0, p_r^0, p_\varphi^0) > 1$, and angular momentum $p_\phi^0$; from these, the initial radial momentum $p_r^0$ can be solved for following~\cite{Nagar:2020xsk,Damour:2014afa}.
Setting $r_0$ to a sufficiently large value (we use $r_0 = 3000M$), as the \acp{BH} draw closer, they lose energy and momentum via \ac{GW} emission, with an especially strong burst of radiation occurring at the time of closest approach.
The dynamical phenomenology is determined by the evolution of the energy: if $E/M$ remains above 1 after the encounter, the \acp{BH} scatter along asymptotically free trajectories; if instead $E/M$ dips below 1, the system becomes bound, an outcome that we refer to as a dynamical capture.
Ref.~\cite{Nagar:2020xsk} explored the $E_0,p_\phi^0$ parameter space, showing that \acp{BBH} undergo a direct plunge if the initial energy is sufficiently high and/or $p_\phi^0$ is below a critical threshold $p_\phi^{\rm LSO}$.\footnote{$p_\phi^{\rm LSO}$ is defined as the angular momentum of the (conservative) last stable (circular) orbit, or LSO. For values $p_\phi^0 < p_\phi^{\rm LSO}$, the effective potential $V(r) = H_{\rm EOB}(r, p_\phi, p_r=0)$ has a peak value smaller than 1, resulting in no radial turning points if $E_0/M > 1$.}
For each mass ratio and spin configuration, a line in the $(E_0, p_\phi^0)$ plane separates scattering events from dynamical captures; bordering this separatrix there is a region where the system becomes bound after the first encounter, but transitions to a highly eccentric orbit, going through a few orbital cycles before merging.

The waveform displays an approximately constant strain as the \acp{BH} approach or recede from each other, when $\dot{\phi} \sim 0$ and $|\dot{r}| \sim \text{constant}$.
On scattering trajectories, each mode's amplitude and frequency peak at closest approach, with $h_{\ell m}$ evolving through more wave cycles around the maximum the lower the energy.
The same is true for dynamical captures, where lower energies mean that the \acp{BH} spend more time circling each other at close separation before merging, leading to a pre-merger morphology that is closer to that of a quasi-circular inspiral.
High-energy systems feature instead a quick growth in the strain amplitude just before merger, while going through few or no wave cycles.
A quieter pre-merger stage is the key feature that we take as a marker for a potential hyperbolic nature for a \ac{GW} event.

The accuracy of the \dali~dynamical model can be evaluated by comparison with \ac{NR}-computed scattering angles~\cite{Nagar:2024oyk,Albanesi:2024xus,Albanesi:2025txj,Long:2025nmj}.
The model performs quite well in the low energy limit, while its accuracy progressively degrades moving towards configurations where the \acp{BH} interact more closely in the strong field.
Ref.~\cite{Albanesi:2024xus} found the model capable of accurately predicting the orbit morphology of \ac{NR} simulations, i.e., whether for given initial data the binary would scatter or merge, at lower energies, including for dynamical captures resulting in eccentric orbits undergoing a handful of close encounters before merger.
However, the \ac{EOB} separatrix between scattering and capture was also found there to diverge from the numerical one as $E_0$ grows.
The effectiveness of the noncircular waveform model in the hyperbolic regime has been investigated in a few works:
Ref.~\cite{Albanesi:2021rby} focused on the extreme mass ratio limit, comparing \dali~with exact waveforms from numerical solutions of the Teukolsky equation; in~\cite{Andrade:2023trh} the authors instead used a small set of \ac{NR} simulations of spinning and non-spinning \acp{BBH}.
In both cases, time-domain phasing comparisons show that the non-circular-corrected TEOB waveform matches the numerical targets to a satisfactory degree (phase differences of $O(0.1 \, {\rm rad})$), with issues mainly arising in the transition from plunge to ringdown in the case of dynamical captures.
These can be traced to two factors: the merger-ringdown model is only informed by, and adapted to, quasi-circular inspirals, while studies including the above works have shown that the post-inspiral waveform is affected by noncircular dynamics~\cite{Carullo:2023kvj,Rao:2026lmz}.
Furthermore, the \ac{nqc} corrections are often unable to enforce a smooth connection of the pre- and post-merger portions of the waveform for direct plunges (or highly eccentric orbits), due to failures in the delicate computation of the coefficients entering the \ac{nqc} factor, resulting in erratic results in the worst cases.
In order to avoid these unphysical features, we do not use \ac{nqc} corrections in the waveform in our analyses.
Refs.~\cite{Andrade:2023trh, Albanesi:2025txj} additionally compute mismatches between \dali~and \ac{NR} hyperbolic waveforms: while, as expected, results are worse for systems that reach deeper into the strong field (low angular momentum, high energy), \ac{EOB}/\ac{NR} mismatches mostly cluster around the 1\% level for dynamical captures, with a few simulations exceeding the 3\% threshold. 

\subsection{GW Hyperbolic Model Parameters}
\label{subsec:params}
For a \ac{BBH} on a \ac{QC} inspiral, the \ac{GW} signal is determined by eight \textit{intrinsic} quantities -- the two masses and six spin vector components $\Omega = \{ m_1, m_2, s_{1,x}, s_{1,y}, s_{1,z}, s_{2,x}, s_{2,y}, s_{2,z} \} $ -- and seven observer-dependent \textit{extrinsic} parameters $\Theta =\{\theta_{\rm JN}, \phi_c, \psi, d_L, ra, dec, t_c\}$, where $\theta_{\rm JN}$ is the inclination relative to the line of sight, $\phi_c$ is the coalescence phase, $\psi$ is the polarization angle, $d_L$ is the luminosity distance, $ra$ and $dec$ are the right ascension and declination angles that define the source's position in the sky, and $t_c$ is the coalescence time.\footnote{
    $t_c$ is usually defined for merging systems as coinciding with the peak of the $(2, \pm 2)$ modes.
    As mentioned in~\cite{Henshaw_2025}, an amended definition is needed for zoom-whirl orbits, where the amplitude can peak at an earlier periastron passage rather than at merger.
    In the case of scattering orbits, $t_c$ is still the peak $(2, \pm 2)$ amplitude time, although of course it corresponds to just the closest approach, rather than the coalescence.
}
To describe a system on a hyperbolic orbit, additional intrinsic parameters are required.
They can be chosen to be $E_0$, the initial energy of the system, and $p_\phi^0$, its initial angular momentum.

We express the masses in terms of solar masses ($M_\odot$), and use the dimensionless spin parameters $\chi_{i,x}, \chi_{i,y}, \chi_{i,z}$.
For precessing systems, these are defined relative to a frame with $\hat{\bm{z}}=\hat{\bm{L}}$, at some reference frequency; we restrict to spins aligned with $\bm{L}$ in the case of hyperbolic orbits, so $\chi_{i, x} = \chi_{i, y} = 0$ at all times.
We use dimensionless measures of the hyperbolic parameters by rescaling with the masses, the initial energy as $E_0/M$, and the initial angular momentum as $p^0_\phi = L/(M \mu)$.

\subsection{RIFT}
\label{subsec:rift}
To perform all PE presented in this work, we use the algorithm~\RIFT~\cite{Lange2018,PhysRevD.99.084026,PhysRevD.107.024040,Wagner:2025bih}.
This highly parallelizable, iterative code esimates the posterior probability distribution $\hat{p}_{\rm post}$ of the binary parameters $(\Theta, \Omega)$ that best describe a \ac{GW} signal via Bayes's Theorem,
\begin{align}
    p_{\rm post}(\Theta, \Omega) \simeq  \frac{{{\cal L}}(\Theta, \Omega) p(\Theta, \Omega)}{\mathcal{Z}}\,,
\end{align}
where $\mathcal{L}$ is the likelihood function, $p(\Theta, \Omega)$ is the prior probability distribution, and the evidence $\mathcal{Z} = \int d\Theta d\Omega {\mathcal{L}} (\Theta, \Omega) p (\Theta, \Omega)$ is a normalization constant.
Given observational data from a network of $N$ detectors,  $d_k$ with $k = 1, \dots, N$, and considering a template waveform model $h_k (\Theta, \Omega)$, it does so by splitting the set of binary parameters in two: we first focus on estimating the posterior distribution of the intrinsic parameters $\Omega$ by marginalizing over the extrinsic parameters $\Theta$.
Assuming a stationary, Gaussian noise model, the likelihood can be expressed as
\begin{align}
    \label{eq:loglikelihood}
    \ln & {\cal L}(\Theta, \Omega) \propto \nonumber \\
    & -\frac{1}{2}\sum_k \bigl\{ \qmstateproduct{h_k(\Theta,\Omega)-d_k}{h_k(\Theta,\Omega)-d_k}_k
    - \qmstateproduct{d_k}{d_k}_k \bigr\}\, ,
\end{align}
where we have omitted normalization constants.
The angled brackets denote weighted inner products, defined by
\begin{align}
    \qmstateproduct{a}{b}_k \equiv \int_{-\infty}^{\infty} 2 df \frac{\tilde{a}(f)^*\tilde{b}(f)}{S_{n,k}(|f|)}\,,
\end{align}
where $S_{n,k}(|f|)$ is the noise power spectral density in the $k$-th detector, $\tilde{a}(f)$ is the Fourier transform of $a(t)$,
${}^*$ denotes complex conjugation,
and $f$ is the frequency.
We adopt a low-frequency cutoff $f_{\rm low}$ such that all inner products are modified to
\begin{eqnarray}
    \qmstateproduct{a}{b}_k\equiv 2 \int_{|f|>f_{\rm low}}  df \frac{\tilde{a}(f)^*\tilde{b}(f)}{S_{n,k}(|f|)} \,.
\end{eqnarray}
In the first stage of the \RIFT algorithm, the \emph{marginal} likelihood is evaluated for a discrete set of intrinsic points $\ln {\cal L}_\alpha \equiv \ln {\cal L}_{\rm marg}(\Omega_\alpha)$ by integrating over the extrinsic prior space,
\begin{align}
    {\cal L}_{\rm marg}(\Omega) \equiv \int d\Theta p(\Theta) {\cal L}(\Omega,\Theta)\,.
\end{align}
In the second stage, an interpolative procedure is applied to the training set $\{(\Omega_\alpha,{\cal L}_\alpha)\}$ to produce a continuous likelihood distribution $\hat{\cal L}(\Omega)$ to be used in Bayes' Theorem to construct the posterior distribution over the intrinsic parameters $\Omega$:
\begin{align}
    {p}_{\rm post}(\Omega) \simeq  \frac{{{\cal L}}(\Omega) p(\Omega)}{\mathcal{Z}}\,,
\end{align}
where $p(\Omega)$ is the prior distribution of the intrinsic parameters.
Via the Monte Carlo integration,~\RIFT provides independent fair draws from this estimated posterior that can then be used as the new starting grid for the next set of marginalized log-likelihood evaluations.
After several iterations, the set of discrete marginalized log-likelihood points yields an interpolated log-likelihood distribution converged to the true $\ln \mathcal{L}_{\rm marg}$ that then yields the true intrinsic posterior distribution.
Once convergence is reached, \RIFT performs the marginalization again on the final intrinsic posterior points, saving the extrinsic data, and resamples to associate intrinsic points to their corresponding extrinsic points.

\subsection{Settings}
\label{subsec:settings}
When defining the priors, we generally follow the standard settings used in the \ac{gwtc} papers~\cite{LIGOScientific:2021usb,KAGRA:2021vkt,LIGOScientific:2025slb}.
For all parameters except spins and luminosity distance, we use a uniform distribution across relevant value ranges\footnote{For inclination, we sample uniformly in $\cos\iota$.}.
For the spins in the non-precessing analyses, we adopt an aligned-spin prior that is equivalent to one uniform in spin magnitude after marginalizing out the extra spin degrees of freedom~\cite{Lange2018}.
As for the precessing analyses, we adopt a distribution that is uniform in the spin magnitudes $|\bm{\chi}_{i}|$ and isotropic in spin orientation, with $\chi_{1,x,y,z},\chi_{2,x,y,z} \in\{-0.99, 0.99\}$.
For the masses, we use a uniform distribution in the detector frame component masses, with bounds defined in terms of total mass and symmetric mass ratio, with event-dependent ranges.
For the luminosity distance, we used a uniform prior that corresponds to uniform in merger rate of comoving volume and time, with varying $d_{L,\rm max}$ depending on the event.

Priors for the hyperbolic parameters $E_0, p_\phi^0$ are both uniform distributions, once again with event-dependent ranges, mostly set by model limitations.
We restrict most events to a maximum energy of $E_0^{\rm max} = 1.2$.
During initial testing, we found that \dali~does not consistently produce a waveform above this threshold due to failures in the dynamical model.
In fact, due to the mass-ratio dependent relationship between the real and effective energies (see Eq.~\eqref{eq:eob_ham}), these failures occur at progressively lower values of $E_0$ as $q$ increases; intuitively, the meaning of ``high energy'' shifts downward for more mass-asymmetric systems.
Because of this, events whose posterior distributions include significant support for unequal mass ratios require a lower upper limit $E_0^{\rm max}$, which we set on a case by case basis; the lowest value we use among the event catalog is $E_0^{\rm max} = 1.08$.
The model is less sensitive to the initial angular momentum; therefore, we use an upper limit in the range $7 < p^{0, {\rm max}}_{\phi}< 25$, depending on the event.
These different $p^{0, {\rm max}}_{\phi}$ are chosen to avoid any railing of the posterior at high values of $p_\phi^0$; i.e., they are increased until the posterior distribution has sufficiently decayed to zero at $p^{0, {\rm max}}_\phi$, so the entire posterior-supported region is contained within the prior range.
For all events, we set $E_0^{\rm min} = 1.002$, slightly above the bound-unbound separatrix, and $p^{0, \rm min}_{\phi} = 2$.

For all analyses, we assume perfect calibration (i.e., we do not marginalize over calibration uncertainty).
While~\RIFT can produce reweighted samples that take this into account~\cite{Wagner:2025bih}, the method it uses (see \cite{2020PhRvD.102l2004P}) requires the waveform model to be available in the standard \texttt{LALSuite} code infrastructure~\cite{lalsuite,WETTE2020100634}.
As the hyperbolic configuration of~\dali~is not available there, we skip this step. However, calibration uncertainties are expected to have little impact on the estimation of the intrinsic parameters at the signal-to-noise ratios we analyze here, primarily being absorbed at leading order into refinement in the event's sky location and event time; see~\cite{Vitale:2011wu,2020PhRvD.102l2004P}.

\section{Results}
\label{sec:results}
In this section, we present the results of our analyses of a total of \totalevents publicly-available \ac{GW} events available on the \ac{LVK}'s \ac{GWOSC}~\cite{2021SoftX..1300658A,LIGOScientific:2019lzm,KAGRA:2023pio,LIGOScientific:2025snk}.
Table \ref{tab:events} shows the list of events analyzed.
This list was chosen such that one of the median component masses fall within the pair-instability mass gap, which we take to be $m_*\ge50 M_\odot$~\cite{2021ApJ...912L..31W}.
For all these events, we carry out \ac{PE} under both the hyperbolic, non-precessing and \ac{QC}, spin-precessing hypotheses, using the \dali~waveform model.
In the particularly interesting cases of GW190521 and GW231123, we also perform analyses with eccentric, non-precessing and eccentric, precessing waveforms from the same \ac{GW} model, as well as the \ac{QC}, spin-precessing NR surrogate model \nrsur~\cite{2019PhRvR...1c3015V}, to build confidence in our conclusions about these two exceptional events.

To quantify the relative change in the estimation of a parameter $x$ between two analyses, we define:
\begin{equation}
    \epsilon_x = \dfrac{|\mu_1 - \mu_2|}{\sqrt{\sigma_1^2+\sigma_2^2}}\, ,
\end{equation}
where $\mu_k,\sigma_k$ are the posterior mean and standard deviation of $x$ obtained using models $1$ (here, hyperbolic) and $2$ (here, precessing). To quantify the relative support for two competing hypotheses $\mathcal{H}_1$ and $\mathcal{H}_2$, we calculate the \ac{BF} defined as
\begin{equation}
    \ln \mathcal{B}^{\mathcal{H}_1}_{\mathcal{H}_2}=\ln\frac{\mathcal{Z}(\mathcal{H}_1)}{\mathcal{Z}(\mathcal{H}_2)},
\end{equation}
where $\mathcal{Z}(\mathcal{H}_k)$ is the evidence associated to $\mathcal{H}_k$.

\subsection{Masses}
\label{subsec:masses}
By analyzing each event with both a hyperbolic and a \ac{QC}, precessing model, we are able to directly compare the inferred \ac{BBH} parameters under the two scenarios.
A summary of these results is plotted for each event in Figure~\ref{fig:violin}.
Regarding the masses, while there are differences between the two analyses, for most events there is at least some overlap between the two posteriors.
We find a mostly systematic shift of the total source mass posterior to lower values when assuming \ac{QC}, precessing systems.
Events with the largest differences in $M_{\rm source}$, as well as in the detector-frame $M$, are: GW170729, GW190413, GW190706, GW191109\_010717, GW191127\_050227, GW230914\_111401, and GW231028\_153006. As for the mass ratio, the two analyses return generally more consistent posterior distributions, although we find marked differences in a few cases, the largest being GW190929, GW200220\_061928, GW230630\_125806, GW230820\_212515, and GW231028\_153006. To quantify these relative changes between two posteriors, we calculate $\epsilon_x$ defined in Section~\ref{subsec:settings}. When doing so for $M_{\rm source}$ and $q$, we find the largest values to be $\epsilon_{M_{\rm source}}=1.578,\epsilon_q=0.708$ for events GW230914\_111401 and GW231028\_153006 respectively.

\subsection{Spins}
\label{subsec:spins}
Similarly to the masses in Section~\ref{subsec:masses}, we can also directly compare the spin posteriors with the non-precessing parameters of the two analyzes, again see Figure~\ref{fig:violin}.
We focus on the effective aligned spin parameter $\chi_{\rm eff}$ (see Eq.~\eqref{eq:chieff}), which is better constrained in \ac{PE} than the individual spin parameters due to it affecting the \ac{GW} signal at lower \ac{pn} order.
We quantify the magnitude of the perpendicular spin components, which drive the precession of the orbital plane, via the effective parameter $\chi_{\rm p}$, defined in Eq.~\eqref{eq:chip}.

For $\chi_{\rm eff}$, results vary of our two analyses returning posteriors ranging from highly consistent to, in at least one case, radically divergent.
Events displaying the largest differences are GW190521, GW1901109\_010717, GW230914\_111401, and GW231028\_153006.
The case of GW1901109\_010717 is particularly notable, as the two posteriors appear almost symmetric about 0 and have no overlap with the largest $\epsilon_{\chi_{\rm eff}}$ value of $\epsilon_{\chi_{\rm eff}}=4.142$.
However, it is hard when evaluating results for $\chi_{\rm eff}$ to disentangle genuine effects of sampling on hyperbolic systems from those due to the lack of spin-precession.
When analyzing a significantly precessing signal with a non-precessing model, the \ac{PE} can compensate the unmodeled spin components by recovering a positive $\chi_{\rm eff}$~\cite{Ng:2018neg} (see $\chi_{\rm p}$ posteriors in Figure~\ref{fig:violin}).
This seems to occur for a few of our events, such as GW190521 and the aforementioned GW191109\_010717, which is again a stark example.
However, this doesn't seem to always be the case; see GW230914\_111401 and GW231028\_153006 for examples.
To further investigate any potential systematic effect on spin recovery when sampling with hyperbolic parameters, a hyperbolic, precessing model is needed to decouple actual effects from missing physics.

\afterpage{\clearpage
\begin{widetext}
\begin{figure*}
    \includegraphics[width=0.9\textwidth]{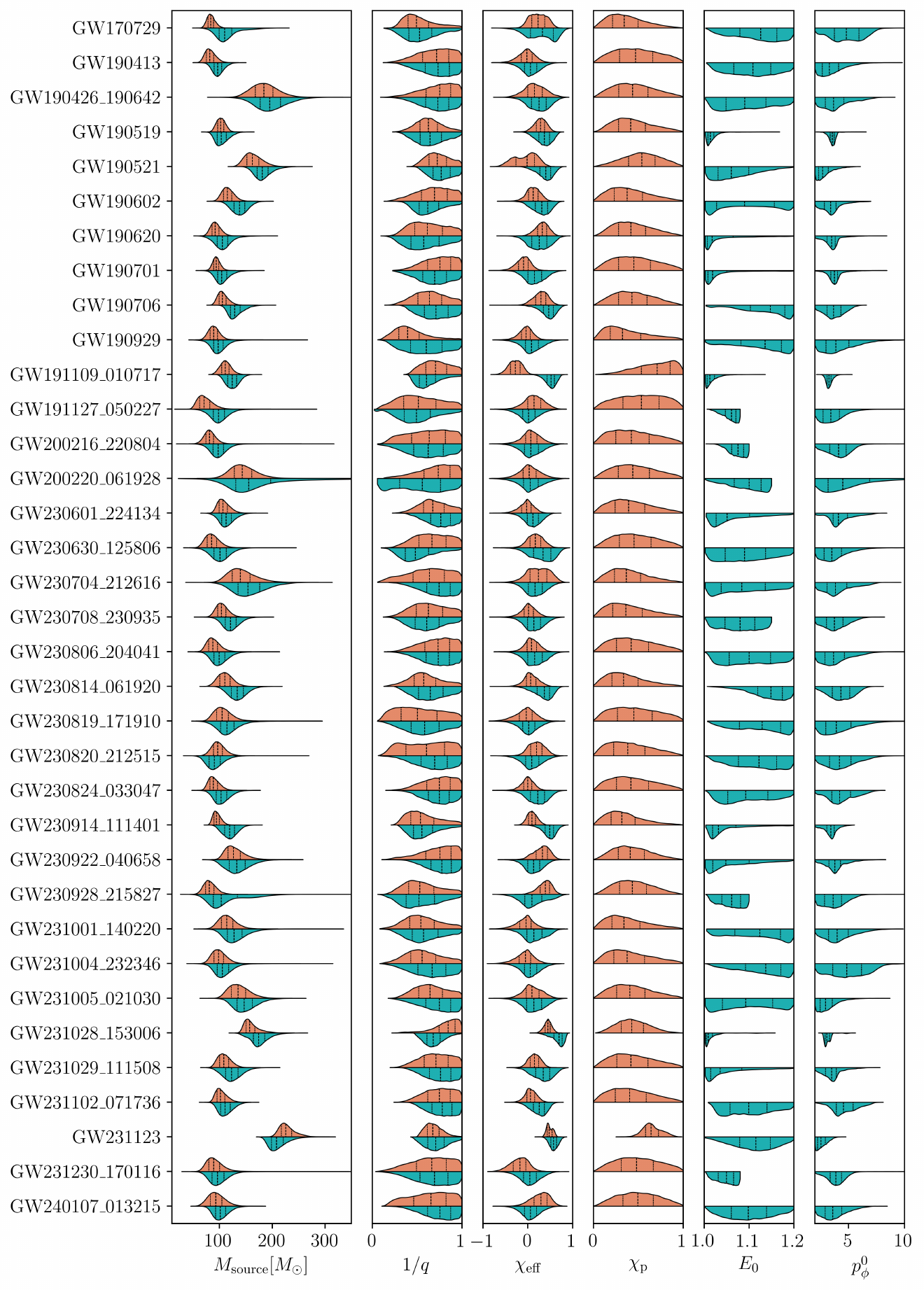}
    
    \caption{
        \label{fig:violin}
        \textbf{Marginal posterior distributions for all events:} This figure shows violin plots comparing the posteriors of the hyperbolic and precessing analyses for all \totalevents~events obtained using \dali.
        The top (orange) and bottom (blue) distributions correspond to the precessing and hyperbolic analyses, respectively.
        The parameters shown are the source frame total mass ($M_{\rm source}$), (inverted) mass ratio ($1/q$), inspiral aligned ($\chi_{\rm eff}$) and precessing ($\chi_{\rm p}$) spin parameters, initial energy ($E_0$), and initial angular momentum ($p_{\phi0}$).
        }
\end{figure*}\clearpage
\end{widetext}
}
\afterpage{\clearpage
\begin{widetext}

\begin{table*}
\centering
\begin{tabular}{l l c c c c c c c}
    \toprule
    \textbf{Event} &  & \bmath{$M_{\mathrm{source}} [M_{\odot}]$} & \bmath{$q$} & \bmath{$\chi_{\mathrm{eff}}$} & \bmath{$\chi_{\mathrm{p}}$} & \bmath{$E_0$} & \bmath{$p_{\varphi}^{0}$} & \bmath{$\ln \mathcal{B}^{\rm hyp}_{\rm prec}$} \\
    \midrule
    \multirow{2}{3.25cm}[-0.5ex]{GW170729          } & Precessing & $83.39_{-11.24}^{+14.65}$ & $0.49_{-0.19}^{+0.40}$ & $0.23_{-0.29}^{+0.30}$ & $0.34_{-0.25}^{+0.40}$ &  &  & \multirow{2}{2cm}[-0.5ex]{\centering $-16.24^{+0.09}_{-0.09}$}  \\[0.4em]
     & Hyperbolic & $109.97_{-24.39}^{+50.77}$ & $0.53_{-0.25}^{+0.36}$ & $0.33_{-0.52}^{+0.41}$ &  & $1.13_{-0.10}^{+0.07}$ & $4.81_{-2.36}^{+2.23}$ &   \\
    \midrule
    \multirow{2}{3.25cm}[-0.5ex]{GW190403\_051519  } & Precessing & $107.26_{-29.78}^{+38.49}$ & $0.35_{-0.22}^{+0.55}$ & $0.62_{-0.54}^{+0.20}$ & $0.38_{-0.26}^{+0.37}$ &  &  & \multirow{2}{2cm}[-0.5ex]{\centering $-4.68^{+0.09}_{-0.09}$}  \\[0.4em]
     & Hyperbolic & $112.67_{-72.89}^{+140.80}$ & $0.40_{-0.34}^{+0.52}$ & $0.02_{-0.46}^{+0.50}$ &  & $1.06_{-0.05}^{+0.04}$ & $9.75_{-7.19}^{+13.65}$ &   \\
    \midrule
    \multirow{2}{3.25cm}[-0.5ex]{GW190413          } & Precessing & $81.63_{-13.76}^{+19.00}$ & $0.72_{-0.35}^{+0.25}$ & $-0.02_{-0.33}^{+0.29}$ & $0.47_{-0.35}^{+0.41}$ &  &  & \multirow{2}{2cm}[-0.5ex]{\centering $-2.04^{+0.07}_{-0.07}$}  \\[0.4em]
     & Hyperbolic & $96.79_{-15.65}^{+19.31}$ & $0.73_{-0.33}^{+0.24}$ & $0.06_{-0.34}^{+0.34}$ &  & $1.11_{-0.08}^{+0.08}$ & $3.29_{-1.15}^{+2.31}$ &   \\
    \midrule
    \multirow{2}{3.25cm}[-0.5ex]{GW190426\_190642  } & Precessing & $184.18_{-37.27}^{+44.25}$ & $0.75_{-0.37}^{+0.22}$ & $0.15_{-0.38}^{+0.40}$ & $0.43_{-0.32}^{+0.40}$ &  &  & \multirow{2}{2cm}[-0.5ex]{\centering $-4.28^{+0.06}_{-0.06}$}  \\[0.4em]
     & Hyperbolic & $195.37_{-42.63}^{+52.70}$ & $0.73_{-0.32}^{+0.23}$ & $0.25_{-0.39}^{+0.35}$ &  & $1.09_{-0.07}^{+0.10}$ & $3.68_{-1.46}^{+2.71}$ &   \\
    \midrule
    \multirow{2}{3.25cm}[-0.5ex]{GW190519          } & Precessing & $102.42_{-12.90}^{+14.08}$ & $0.62_{-0.22}^{+0.27}$ & $0.28_{-0.24}^{+0.21}$ & $0.41_{-0.28}^{+0.42}$ &  &  & \multirow{2}{2cm}[-0.5ex]{\centering $-11.26^{+0.14}_{-0.14}$}  \\[0.4em]
     & Hyperbolic & $103.44_{-15.99}^{+23.87}$ & $0.65_{-0.24}^{+0.28}$ & $0.37_{-0.30}^{+0.24}$ &  & $1.01_{-0.01}^{+0.03}$ & $3.56_{-0.54}^{+0.39}$ &   \\
    \midrule
    \multirow{2}{3.25cm}[-0.5ex]{GW190521          } & Precessing & $162.27_{-22.09}^{+31.54}$ & $0.72_{-0.19}^{+0.23}$ & $-0.02_{-0.47}^{+0.34}$ & $0.54_{-0.35}^{+0.33}$ &  &  & \multirow{2}{2cm}[-0.5ex]{\centering $3.71^{+0.11}_{-0.11}$}  \\[0.4em]
     & Hyperbolic & $181.03_{-21.97}^{+32.82}$ & $0.77_{-0.19}^{+0.20}$ & $0.41_{-0.32}^{+0.24}$ &  & $1.06_{-0.05}^{+0.09}$ & $2.68_{-0.60}^{+1.61}$ &   \\
    \midrule
    \multirow{2}{3.25cm}[-0.5ex]{GW190602          } & Precessing & $115.10_{-16.17}^{+20.39}$ & $0.69_{-0.34}^{+0.27}$ & $0.12_{-0.26}^{+0.24}$ & $0.37_{-0.28}^{+0.43}$ &  &  & \multirow{2}{2cm}[-0.5ex]{\centering $-7.73^{+0.07}_{-0.07}$}  \\[0.4em]
     & Hyperbolic & $137.94_{-23.63}^{+24.71}$ & $0.62_{-0.20}^{+0.31}$ & $0.31_{-0.34}^{+0.28}$ &  & $1.09_{-0.08}^{+0.10}$ & $3.41_{-1.21}^{+1.89}$ &   \\
    \midrule
    \multirow{2}{3.25cm}[-0.5ex]{GW190620          } & Precessing & $92.05_{-13.83}^{+17.77}$ & $0.61_{-0.28}^{+0.34}$ & $0.32_{-0.26}^{+0.22}$ & $0.42_{-0.28}^{+0.39}$ &  &  & \multirow{2}{2cm}[-0.5ex]{\centering $-19.38^{+0.13}_{-0.13}$}  \\[0.4em]
     & Hyperbolic & $105.59_{-24.50}^{+29.47}$ & $0.57_{-0.29}^{+0.36}$ & $0.26_{-0.37}^{+0.36}$ &  & $1.02_{-0.02}^{+0.14}$ & $3.51_{-1.19}^{+1.31}$ &   \\
    \midrule
    \multirow{2}{3.25cm}[-0.5ex]{GW190701          } & Precessing & $93.80_{-10.74}^{+12.78}$ & $0.75_{-0.31}^{+0.22}$ & $-0.09_{-0.29}^{+0.24}$ & $0.45_{-0.33}^{+0.42}$ &  &  & \multirow{2}{2cm}[-0.5ex]{\centering $-10.66^{+0.12}_{-0.12}$}  \\[0.4em]
     & Hyperbolic & $104.31_{-18.21}^{+27.04}$ & $0.70_{-0.31}^{+0.27}$ & $0.15_{-0.34}^{+0.36}$ &  & $1.02_{-0.02}^{+0.15}$ & $3.74_{-0.94}^{+1.44}$ &   \\
    \midrule
    \multirow{2}{3.25cm}[-0.5ex]{GW190706          } & Precessing & $105.33_{-13.23}^{+19.98}$ & $0.64_{-0.29}^{+0.31}$ & $0.30_{-0.29}^{+0.24}$ & $0.43_{-0.30}^{+0.39}$ &  &  & \multirow{2}{2cm}[-0.5ex]{\centering $-7.16^{+0.10}_{-0.10}$}  \\[0.4em]
     & Hyperbolic & $128.78_{-16.79}^{+27.79}$ & $0.71_{-0.27}^{+0.26}$ & $0.47_{-0.33}^{+0.22}$ &  & $1.15_{-0.11}^{+0.05}$ & $3.68_{-1.38}^{+1.47}$ &   \\
    \midrule
    \multirow{2}{3.25cm}[-0.5ex]{GW190929          } & Precessing & $89.01_{-15.60}^{+19.23}$ & $0.39_{-0.21}^{+0.41}$ & $-0.04_{-0.29}^{+0.24}$ & $0.32_{-0.25}^{+0.51}$ &  &  & \multirow{2}{2cm}[-0.5ex]{\centering $-8.08^{+0.06}_{-0.06}$}  \\[0.4em]
     & Hyperbolic & $97.59_{-21.05}^{+26.73}$ & $0.60_{-0.35}^{+0.35}$ & $0.08_{-0.37}^{+0.38}$ &  & $1.14_{-0.12}^{+0.06}$ & $3.87_{-1.63}^{+3.03}$ &   \\
    \midrule
    \multirow{2}{3.25cm}[-0.5ex]{GW191109\_010717  } & Precessing & $111.10_{-16.29}^{+18.38}$ & $0.71_{-0.23}^{+0.25}$ & $-0.28_{-0.25}^{+0.23}$ & $0.71_{-0.41}^{+0.25}$ &  &  & \multirow{2}{2cm}[-0.5ex]{\centering $-6.79^{+0.15}_{-0.15}$}  \\[0.4em]
     & Hyperbolic & $124.22_{-18.11}^{+19.91}$ & $0.60_{-0.15}^{+0.29}$ & $0.52_{-0.23}^{+0.17}$ &  & $1.01_{-0.01}^{+0.03}$ & $3.26_{-0.37}^{+0.56}$ &   \\
    \midrule
    \multirow{2}{3.25cm}[-0.5ex]{GW191127\_050227  } & Precessing & $70.93_{-17.45}^{+28.37}$ & $0.51_{-0.34}^{+0.42}$ & $0.14_{-0.36}^{+0.36}$ & $0.53_{-0.41}^{+0.38}$ &  &  & \multirow{2}{2cm}[-0.5ex]{\centering $-8.77^{+0.09}_{-0.09}$}  \\[0.4em]
     & Hyperbolic & $98.43_{-25.72}^{+29.57}$ & $0.49_{-0.30}^{+0.40}$ & $0.02_{-0.36}^{+0.37}$ &  & $1.06_{-0.03}^{+0.02}$ & $3.42_{-1.27}^{+2.33}$ &   \\
    \midrule
    \multirow{2}{3.25cm}[-0.5ex]{GW200208\_222617  } & Precessing & $52.91_{-17.29}^{+69.87}$ & $0.32_{-0.23}^{+0.58}$ & $0.36_{-0.47}^{+0.42}$ & $0.40_{-0.30}^{+0.42}$ &  &  & \multirow{2}{2cm}[-0.5ex]{\centering $-2.08^{+0.08}_{-0.08}$}  \\[0.4em]
     & Hyperbolic & $210.81_{-147.39}^{+224.61}$ & $0.37_{-0.31}^{+0.54}$ & $-0.00_{-0.46}^{+0.45}$ &  & $1.05_{-0.04}^{+0.04}$ & $13.33_{-10.14}^{+10.51}$ &   \\
    \midrule
    \multirow{2}{3.25cm}[-0.5ex]{GW200216\_220804  } & Precessing & $81.16_{-16.34}^{+19.85}$ & $0.63_{-0.40}^{+0.33}$ & $0.07_{-0.33}^{+0.33}$ & $0.42_{-0.33}^{+0.44}$ &  &  & \multirow{2}{2cm}[-0.5ex]{\centering $-4.91^{+0.10}_{-0.10}$}  \\[0.4em]
     & Hyperbolic & $97.39_{-23.24}^{+28.48}$ & $0.62_{-0.38}^{+0.33}$ & $0.07_{-0.38}^{+0.41}$ &  & $1.08_{-0.03}^{+0.02}$ & $4.12_{-1.76}^{+1.65}$ &   \\
    \midrule
    \multirow{2}{3.25cm}[-0.5ex]{GW200220\_061928  } & Precessing & $143.31_{-34.57}^{+51.53}$ & $0.74_{-0.41}^{+0.24}$ & $0.03_{-0.37}^{+0.40}$ & $0.44_{-0.33}^{+0.42}$ &  &  & \multirow{2}{2cm}[-0.5ex]{\centering $-5.15^{+0.06}_{-0.06}$}  \\[0.4em]
     & Hyperbolic & $155.56_{-57.18}^{+174.85}$ & $0.56_{-0.45}^{+0.39}$ & $0.06_{-0.43}^{+0.45}$ &  & $1.10_{-0.08}^{+0.05}$ & $4.49_{-2.25}^{+13.54}$ &   \\
    \midrule
    \multirow{2}{3.25cm}[-0.5ex]{GW200308\_173609  } & Precessing & $51.90_{-14.36}^{+59.02}$ & $0.36_{-0.24}^{+0.49}$ & $0.57_{-0.68}^{+0.21}$ & $0.40_{-0.27}^{+0.35}$ &  &  & \multirow{2}{2cm}[-0.5ex]{\centering $-1.70^{+0.10}_{-0.10}$}  \\[0.4em]
     & Hyperbolic & $102.24_{-64.25}^{+78.05}$ & $0.39_{-0.32}^{+0.52}$ & $-0.00_{-0.46}^{+0.45}$ &  & $1.06_{-0.05}^{+0.04}$ & $7.30_{-4.91}^{+15.77}$ &   \\
    \midrule
    \multirow{2}{3.25cm}[-0.5ex]{GW230601\_224134  } & Precessing & $106.54_{-14.58}^{+22.08}$ & $0.67_{-0.25}^{+0.28}$ & $-0.02_{-0.31}^{+0.25}$ & $0.39_{-0.30}^{+0.43}$ &  &  & \multirow{2}{2cm}[-0.5ex]{\centering $-10.50^{+0.11}_{-0.11}$}  \\[0.4em]
     & Hyperbolic & $112.71_{-20.77}^{+26.13}$ & $0.76_{-0.26}^{+0.21}$ & $0.11_{-0.32}^{+0.29}$ &  & $1.05_{-0.04}^{+0.12}$ & $4.10_{-1.52}^{+2.40}$ &   \\
    \midrule
    \multirow{2}{3.25cm}[-0.5ex]{GW230630\_125806  } & Precessing & $85.20_{-18.64}^{+25.01}$ & $0.67_{-0.31}^{+0.29}$ & $0.17_{-0.28}^{+0.29}$ & $0.45_{-0.33}^{+0.41}$ &  &  & \multirow{2}{2cm}[-0.5ex]{\centering $-7.84^{+0.07}_{-0.07}$}  \\[0.4em]
     & Hyperbolic & $101.05_{-27.87}^{+30.50}$ & $0.48_{-0.23}^{+0.40}$ & $0.34_{-0.46}^{+0.34}$ &  & $1.09_{-0.08}^{+0.10}$ & $3.53_{-1.34}^{+2.46}$ &   \\
    \midrule
    \multirow{2}{3.25cm}[-0.5ex]{GW230704\_212616  } & Precessing & $139.61_{-33.00}^{+49.21}$ & $0.64_{-0.43}^{+0.32}$ & $0.24_{-0.41}^{+0.38}$ & $0.36_{-0.27}^{+0.42}$ &  &  & \multirow{2}{2cm}[-0.5ex]{\centering $-3.26^{+0.06}_{-0.06}$}  \\[0.4em]
     & Hyperbolic & $154.04_{-41.81}^{+57.60}$ & $0.74_{-0.33}^{+0.23}$ & $0.16_{-0.35}^{+0.37}$ &  & $1.08_{-0.07}^{+0.10}$ & $3.84_{-1.58}^{+2.79}$ &   \\
    \midrule
    \multirow{2}{3.25cm}[-0.5ex]{GW230708\_230935  } & Precessing & $103.94_{-16.64}^{+21.47}$ & $0.63_{-0.30}^{+0.33}$ & $0.02_{-0.28}^{+0.27}$ & $0.36_{-0.28}^{+0.46}$ &  &  & \multirow{2}{2cm}[-0.5ex]{\centering $-9.94^{+0.08}_{-0.08}$}  \\[0.4em]
     & Hyperbolic & $120.96_{-23.03}^{+27.05}$ & $0.61_{-0.27}^{+0.32}$ & $0.14_{-0.35}^{+0.35}$ &  & $1.08_{-0.06}^{+0.06}$ & $3.74_{-1.51}^{+1.86}$ &   \\
    \bottomrule
\end{tabular}

\end{table*}

\begin{table*}
\centering
\begin{tabular}{l l c c c c c c c}
    \toprule
    \textbf{Event} &  & \bmath{$M_{\mathrm{source}} [M_{\odot}]$} & \bmath{$q$} & \bmath{$\chi_{\mathrm{eff}}$} & \bmath{$\chi_{\mathrm{p}}$} & \bmath{$E_0$} & \bmath{$p_{\varphi}^{0}$} & \bmath{$\ln \mathcal{B}^{\rm hyp}_{\rm prec}$} \\
    \midrule
    \multirow{2}{3.25cm}[-0.5ex]{GW230806\_204041  } & Precessing & $87.36_{-16.67}^{+24.49}$ & $0.73_{-0.34}^{+0.24}$ & $0.08_{-0.26}^{+0.26}$ & $0.42_{-0.32}^{+0.44}$ &  &  & \multirow{2}{2cm}[-0.5ex]{\centering $-6.91^{+0.06}_{-0.06}$}  \\[0.4em]
     & Hyperbolic & $99.29_{-21.89}^{+29.07}$ & $0.74_{-0.32}^{+0.23}$ & $0.15_{-0.36}^{+0.37}$ &  & $1.10_{-0.08}^{+0.09}$ & $3.67_{-1.47}^{+2.66}$ &   \\
    \midrule
    \multirow{2}{3.25cm}[-0.5ex]{GW230814\_061920  } & Precessing & $110.12_{-21.13}^{+25.38}$ & $0.57_{-0.27}^{+0.34}$ & $0.06_{-0.26}^{+0.29}$ & $0.33_{-0.25}^{+0.44}$ &  &  & \multirow{2}{2cm}[-0.5ex]{\centering $-2.24^{+0.10}_{-0.10}$}  \\[0.4em]
     & Hyperbolic & $133.68_{-25.45}^{+28.92}$ & $0.65_{-0.27}^{+0.30}$ & $0.37_{-0.40}^{+0.25}$ &  & $1.15_{-0.08}^{+0.05}$ & $4.33_{-1.74}^{+1.67}$ &   \\
    \midrule
    \multirow{2}{3.25cm}[-0.5ex]{GW230819\_171910  } & Precessing & $105.05_{-22.15}^{+32.96}$ & $0.50_{-0.34}^{+0.43}$ & $-0.03_{-0.42}^{+0.32}$ & $0.45_{-0.35}^{+0.44}$ &  &  & \multirow{2}{2cm}[-0.5ex]{\centering $-0.13^{+0.07}_{-0.07}$}  \\[0.4em]
     & Hyperbolic & $114.03_{-29.74}^{+40.29}$ & $0.58_{-0.34}^{+0.36}$ & $0.02_{-0.36}^{+0.36}$ &  & $1.13_{-0.10}^{+0.07}$ & $3.91_{-1.69}^{+4.97}$ &   \\
    \midrule
    \multirow{2}{3.25cm}[-0.5ex]{GW230820\_212515  } & Precessing & $96.83_{-18.58}^{+23.49}$ & $0.60_{-0.40}^{+0.35}$ & $0.19_{-0.28}^{+0.29}$ & $0.38_{-0.30}^{+0.44}$ &  &  & \multirow{2}{2cm}[-0.5ex]{\centering $-7.65^{+0.08}_{-0.08}$}  \\[0.4em]
     & Hyperbolic & $90.12_{-21.66}^{+30.31}$ & $0.70_{-0.35}^{+0.27}$ & $0.09_{-0.36}^{+0.39}$ &  & $1.12_{-0.09}^{+0.07}$ & $4.20_{-1.85}^{+2.69}$ &   \\
    \midrule
    \multirow{2}{3.25cm}[-0.5ex]{GW230824\_033047  } & Precessing & $88.40_{-13.91}^{+20.58}$ & $0.75_{-0.32}^{+0.22}$ & $-0.00_{-0.25}^{+0.22}$ & $0.42_{-0.32}^{+0.44}$ &  &  & \multirow{2}{2cm}[-0.5ex]{\centering $-9.97^{+0.09}_{-0.09}$}  \\[0.4em]
     & Hyperbolic & $103.08_{-20.75}^{+27.09}$ & $0.75_{-0.28}^{+0.22}$ & $0.22_{-0.37}^{+0.33}$ &  & $1.09_{-0.07}^{+0.09}$ & $4.22_{-1.75}^{+2.28}$ &   \\
    \midrule
    \multirow{2}{3.25cm}[-0.5ex]{GW230914\_111401  } & Precessing & $94.38_{-9.98}^{+14.68}$ & $0.54_{-0.21}^{+0.38}$ & $0.09_{-0.17}^{+0.20}$ & $0.31_{-0.24}^{+0.42}$ &  &  & \multirow{2}{2cm}[-0.5ex]{\centering $-19.58^{+0.12}_{-0.12}$}  \\[0.4em]
     & Hyperbolic & $119.33_{-21.15}^{+22.59}$ & $0.55_{-0.19}^{+0.33}$ & $0.49_{-0.28}^{+0.20}$ &  & $1.03_{-0.02}^{+0.13}$ & $3.43_{-1.13}^{+0.96}$ &   \\
    \midrule
    \multirow{2}{3.25cm}[-0.5ex]{GW230922\_040658  } & Precessing & $126.51_{-21.53}^{+35.75}$ & $0.75_{-0.41}^{+0.22}$ & $0.33_{-0.33}^{+0.25}$ & $0.41_{-0.28}^{+0.38}$ &  &  & \multirow{2}{2cm}[-0.5ex]{\centering $-8.40^{+0.13}_{-0.13}$}  \\[0.4em]
     & Hyperbolic & $132.79_{-31.37}^{+41.63}$ & $0.78_{-0.28}^{+0.20}$ & $0.13_{-0.31}^{+0.32}$ &  & $1.05_{-0.04}^{+0.12}$ & $3.78_{-1.40}^{+1.91}$ &   \\
    \midrule
    \multirow{2}{3.25cm}[-0.5ex]{GW230928\_215827  } & Precessing & $80.89_{-16.60}^{+22.59}$ & $0.53_{-0.26}^{+0.39}$ & $0.39_{-0.30}^{+0.21}$ & $0.43_{-0.29}^{+0.37}$ &  &  & \multirow{2}{2cm}[-0.5ex]{\centering $-12.71^{+0.10}_{-0.10}$}  \\[0.4em]
     & Hyperbolic & $103.25_{-31.91}^{+82.09}$ & $0.52_{-0.29}^{+0.38}$ & $0.09_{-0.40}^{+0.52}$ &  & $1.06_{-0.04}^{+0.03}$ & $3.64_{-1.46}^{+1.86}$ &   \\
    \midrule
    \multirow{2}{3.25cm}[-0.5ex]{GW231001\_140220  } & Precessing & $113.89_{-22.79}^{+30.64}$ & $0.54_{-0.28}^{+0.38}$ & $-0.04_{-0.38}^{+0.32}$ & $0.34_{-0.26}^{+0.46}$ &  &  & \multirow{2}{2cm}[-0.5ex]{\centering $-4.17^{+0.06}_{-0.06}$}  \\[0.4em]
     & Hyperbolic & $127.78_{-28.00}^{+42.49}$ & $0.57_{-0.28}^{+0.36}$ & $0.14_{-0.38}^{+0.40}$ &  & $1.12_{-0.10}^{+0.07}$ & $4.00_{-1.67}^{+2.68}$ &   \\
    \midrule
    \multirow{2}{3.25cm}[-0.5ex]{GW231004\_232346  } & Precessing & $98.21_{-19.33}^{+25.64}$ & $0.55_{-0.29}^{+0.37}$ & $-0.05_{-0.39}^{+0.30}$ & $0.37_{-0.28}^{+0.47}$ &  &  & \multirow{2}{2cm}[-0.5ex]{\centering $-3.95^{+0.08}_{-0.08}$}  \\[0.4em]
     & Hyperbolic & $105.61_{-22.68}^{+32.14}$ & $0.67_{-0.36}^{+0.29}$ & $0.05_{-0.34}^{+0.36}$ &  & $1.14_{-0.10}^{+0.06}$ & $4.85_{-2.49}^{+2.68}$ &   \\
    \midrule
    \multirow{2}{3.25cm}[-0.5ex]{GW231005\_021030  } & Precessing & $135.12_{-26.67}^{+37.37}$ & $0.64_{-0.27}^{+0.31}$ & $0.09_{-0.33}^{+0.37}$ & $0.40_{-0.30}^{+0.42}$ &  &  & \multirow{2}{2cm}[-0.5ex]{\centering $-2.08^{+0.06}_{-0.06}$}  \\[0.4em]
     & Hyperbolic & $146.77_{-29.20}^{+40.43}$ & $0.75_{-0.35}^{+0.23}$ & $0.13_{-0.38}^{+0.40}$ &  & $1.09_{-0.08}^{+0.10}$ & $2.97_{-0.87}^{+1.77}$ &   \\
    \midrule
    \multirow{2}{3.25cm}[-0.5ex]{GW231028\_153006  } & Precessing & $157.21_{-13.90}^{+31.55}$ & $0.85_{-0.34}^{+0.14}$ & $0.45_{-0.15}^{+0.14}$ & $0.43_{-0.26}^{+0.31}$ &  &  & \multirow{2}{2cm}[-0.5ex]{\centering $-27.43^{+0.20}_{-0.20}$}  \\[0.4em]
     & Hyperbolic & $174.32_{-26.82}^{+32.72}$ & $0.68_{-0.15}^{+0.20}$ & $0.70_{-0.31}^{+0.14}$ &  & $1.01_{-0.01}^{+0.06}$ & $3.08_{-0.28}^{+0.76}$ &   \\
    \midrule
    \multirow{2}{3.25cm}[-0.5ex]{GW231029\_111508  } & Precessing & $108.14_{-17.29}^{+22.95}$ & $0.71_{-0.30}^{+0.26}$ & $0.15_{-0.22}^{+0.22}$ & $0.42_{-0.31}^{+0.44}$ &  &  & \multirow{2}{2cm}[-0.5ex]{\centering $-12.78^{+0.12}_{-0.12}$}  \\[0.4em]
     & Hyperbolic & $122.93_{-26.64}^{+30.45}$ & $0.76_{-0.28}^{+0.21}$ & $0.35_{-0.38}^{+0.32}$ &  & $1.04_{-0.03}^{+0.13}$ & $3.50_{-1.21}^{+1.70}$ &   \\
    \midrule
    \multirow{2}{3.25cm}[-0.5ex]{GW231102\_071736  } & Precessing & $101.98_{-13.32}^{+21.58}$ & $0.75_{-0.29}^{+0.22}$ & $0.06_{-0.21}^{+0.22}$ & $0.40_{-0.31}^{+0.43}$ &  &  & \multirow{2}{2cm}[-0.5ex]{\centering $-12.97^{+0.12}_{-0.12}$}  \\[0.4em]
     & Hyperbolic & $110.57_{-17.87}^{+24.94}$ & $0.78_{-0.25}^{+0.19}$ & $0.26_{-0.32}^{+0.24}$ &  & $1.10_{-0.07}^{+0.09}$ & $4.54_{-1.51}^{+2.04}$ &   \\
    \midrule
    \multirow{2}{3.25cm}[-0.5ex]{GW231123          } & Precessing & $225.75_{-18.17}^{+32.57}$ & $0.67_{-0.13}^{+0.18}$ & $0.48_{-0.09}^{+0.13}$ & $0.64_{-0.15}^{+0.20}$ &  &  & \multirow{2}{2cm}[-0.5ex]{\centering $-15.80^{+0.23}_{-0.23}$}  \\[0.4em]
     & Hyperbolic & $208.13_{-19.51}^{+35.12}$ & $0.71_{-0.19}^{+0.23}$ & $0.58_{-0.15}^{+0.15}$ &  & $1.12_{-0.08}^{+0.07}$ & $2.55_{-0.50}^{+0.92}$ &   \\
    \midrule
    \multirow{2}{3.25cm}[-0.5ex]{GW231230\_170116  } & Precessing & $89.09_{-19.50}^{+33.48}$ & $0.66_{-0.39}^{+0.30}$ & $-0.18_{-0.36}^{+0.30}$ & $0.48_{-0.35}^{+0.41}$ &  &  & \multirow{2}{2cm}[-0.5ex]{\centering $-2.54^{+0.12}_{-0.12}$}  \\[0.4em]
     & Hyperbolic & $96.41_{-23.17}^{+32.38}$ & $0.69_{-0.37}^{+0.27}$ & $0.05_{-0.37}^{+0.37}$ &  & $1.05_{-0.04}^{+0.03}$ & $3.88_{-1.39}^{+1.22}$ &   \\
    \midrule
    \multirow{2}{3.25cm}[-0.5ex]{GW240107\_013215  } & Precessing & $92.94_{-22.66}^{+29.83}$ & $0.65_{-0.40}^{+0.31}$ & $0.30_{-0.38}^{+0.28}$ & $0.49_{-0.35}^{+0.37}$ &  &  & \multirow{2}{2cm}[-0.5ex]{\centering $-0.96^{+0.09}_{-0.09}$}  \\[0.4em]
     & Hyperbolic & $101.48_{-19.12}^{+26.60}$ & $0.75_{-0.32}^{+0.22}$ & $0.06_{-0.33}^{+0.35}$ &  & $1.10_{-0.08}^{+0.08}$ & $3.59_{-1.36}^{+2.14}$ &   \\
    \bottomrule
\end{tabular}
\caption{\label{tab:events}\textbf{Recovered \ac{BBH} parameters for all events:} This table shares a list of all events analyzed with the \ac{QC}, precessing and hyperbolic, non-precessing versions of~\dali.
We report the median values and the 90\% \acp{ci} of each of the source-frame total mass $M$, (inverted) mass ratio $1/q$, inspiral effective spin $\chi_{\rm eff}$, effective precessing spin parameter $\chi_{\rm p}$, initial enegery $E_0$, initial angular momentum $p_{\phi,0}$.
The last column contains the \ac{BF} between the hyperbolic and the precessing hypotheses, $\ln \mathcal{B}^{\rm hyp}_{\rm prec}$. 
Most events analyzed vary between strongly to slightly favoring the \ac{QC}, precessing description.
Only GW190521 stands out as preferring the hyperbolic hypothesis by about 40:1 odds.}

\end{table*}
\clearpage
\end{widetext}
}

\subsection{Initial energy and angular momentum}
\label{subsec:hyp-parms}

The additional intrinsic parameters characterizing a hyperbolic orbit are the initial energy $E_0$ and the initial angular momentum $p_\phi^0$.
These determine the type of hyperbolic system, yielding dramatically different waveform morphologies across the parameter space (i.e. direct plunges, fly-by orbits before capture, or scatters)~\cite{Nagar:2020xsk,Henshaw_2025}.
We show results for $E_0$ and $p_\phi^0$ in the 1D marginalized distributions in ƒtwo columns of Figure~\ref{fig:violin} as well as in the 2D posteriors for all events, while highlighting GW190521 and GW231123, in Figure~\ref{fig:hyp-params}; medians and 90\% \ac{ci} are in Table~\ref{tab:events}.
For the initial energy, we recover broad posteriors similar to the assumed prior for most events.
A few events stand out for returning more informative $E_0$ distributions, such as GW190519, GW190620, GW190701, GW191109\_010717,  GW230601\_224134, GW230914\_111401, GW230922\_040658, GW231028\_153006, and GW231029\_111508, which favor lower energies, many of these with posterior support only in a tight range above $E_0 = 1$.

A subset of events instead favor higher energies; these include: GW190706, GW190929, GW191127\_050227, GW200216\_220804, GW200220\_061928, GW230814\_061920, GW230928\_215827, GW231004\_232346, and GW231230\_170116.
It has been shown the that hyperbolic configuration of~\dali~is less reliable at the high energy/low angular momentum part of parameter space~\cite{Albanesi:2024xus,Albanesi:2025txj}.
As the energy increases, the character of a dynamical capture waveform evolves, from the system going through a few orbits before merger until there is only a direct plunge.
In these high-$E_0$ cases, the waveform features a strain amplitude that grows very quickly from a constant value just before merger, followed by the ringdown, which in \dali~is modeled in the same way as for bound, \ac{QC} inspirals.
Thus, in the frequency domain, these signals are dominated by their post-merger phase, even more so as all of the events we analyze feature large total masses.
Low energy waveforms being a bad match for the inspiral phase of these events might explain why their posteriors support only the higher $E_0$ values, where effectively there is no inspiral, but only a reasonably well-matching ringdown.
New analyses should be carried out to better understand these results once a ringdown model incorporating information from \ac{NR} hyperbolic waveforms is implemented in~\dali.

\begin{table*}
\begin{tabular}{ l c c c c c}
\toprule
    \textbf{Event} &
    \bmath{$\ln \mathcal{B}^{\rm hyp}_{\rm prec}$} &
    \bmath{$\ln \mathcal{B}^{\rm hyp}_{\rm align}$} &
    \bmath{$\ln \mathcal{B}^{\rm hyp}_{\rm ecc}$} &
    \bmath{$\ln \mathcal{B}^{\rm hyp}_{\rm e+p}$} &
    \bmath{$\ln \mathcal{B}^{\rm hyp}_{\rm NRSur}$} \\
\midrule \\[-0.75em]
    GW190521 & $3.71^{+0.11}_{-0.11}$ & $5.01^{+0.13}_{-0.13}$ & $5.64^{+0.13}_{-0.13}$ & $4.18^{+0.13}_{-0.13}$ & $2.01^{+0.11}_{-0.11}$ \\[0.5em]
    GW231123 & $-15.80^{+0.24}_{-0.24}$ & $5.94^{+0.20}_{-0.20}$ & $6.13^{+0.22}_{-0.22}$ & $-12.24^{+0.26}_{-0.26}$ & $-10.14^{+0.24}_{-0.24}$ \\[0.5em]
    GW190521-like, hyp  & $1.83^{+0.12}_{-0.12}$ & - & - & - & - \\[0.5em]
    GW190521-like, prec & $-0.75^{+0.12}_{-0.12}$ & - & - & - & - \\[0.5em]
    GW231123-like, hyp  &  $13.14^{+0.13}_{-0.13}$ & - & - & - & - \\[0.5em]
    GW231123-like, prec & $-21.73^{+0.24}_{-0.24}$ & - & - & - & - \\[0.25em]
\bottomrule
\end{tabular}
\caption{\label{tab:small}%
\textbf{Bayes' factors comparing hyperbolic to other scenarios:} \acp{BF} for the real GW190521 and GW231123 events, and the corresponding synthetic hyperbolic and precessing injections, comparing support for the hyperbolic scenario against the \ac{QC}, precessing (\bmath{$\ln \mathcal{B}^{\rm hyp}_{\rm prec}$}); \ac{QC}, non-precessing (\bmath{$\ln \mathcal{B}^{\rm hyp}_{\rm align}$}); eccentric, non-precessing (\bmath{$\ln \mathcal{B}^{\rm hyp}_{\rm ecc}$}); and eccentric, precessing (\bmath{$\ln \mathcal{B}^{\rm hyp}_{\rm e+p}$}) ones, all using the \dali~model.
We also report the \ac{BF} contrasting our hyperbolic \dali~analysis with the \ac{QC}, precessing one using \nrsur~ (\bmath{$\ln \mathcal{B}^{\rm hyp}_{\rm NRSur}$}).
While GW190521 and GW231123 share some similarities as high-mass, ringdown-dominated events showing support for precessing spins, we find that GW231123 strongly favors the \ac{QC}, precessing scenario, while for GW190521 a hyperbolic description better fits the data.
As for the simulated analyses, the two GW231123-like signals strongly favor their corresponding scenario, while the GW190521-like signals do so only marginally.
}
\end{table*}

\subsection{GW190521}
\label{subsec:gw190521}
\begin{figure*}
    \includegraphics[width=0.85\textwidth]{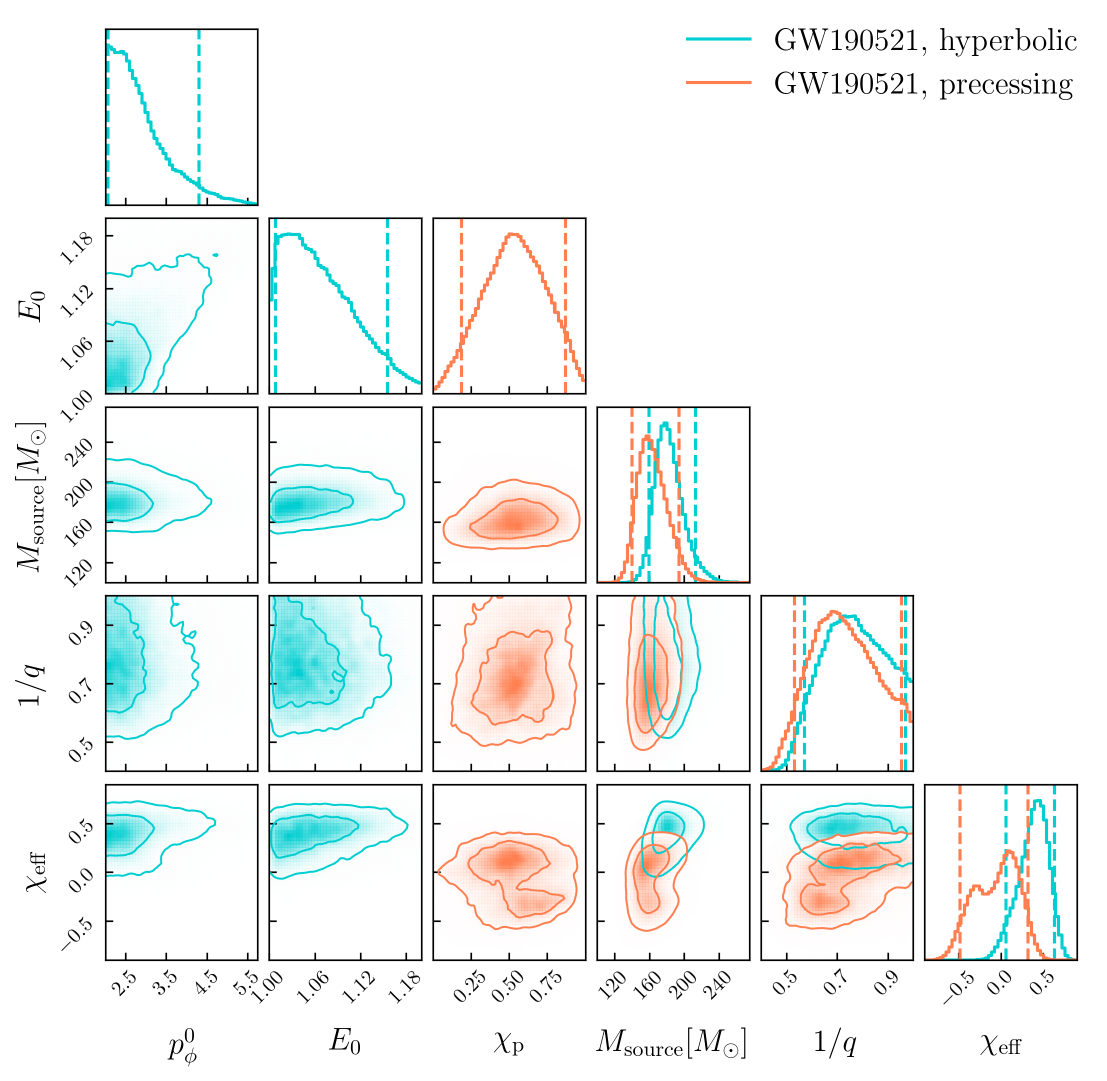}
    
    \caption{
        \label{fig:gw190521}
        \textbf{Hyperbolic and precessing analyses for GW190521:} This figure compares the results of the hyperbolic (teal) and \ac{QC}, precessing (orange) analyses for GW190521 using~\dali.
        The panels at the top of each column are the 1D marginalized posterior distributions for each parameter, while the others show the 2D posteriors and the contours of the $50\%$ and $90\%$ credible regions.
        The parameters shown are the mass ratio $q$, source-frame total mass $M_{\rm source}$, effective aligned ($\chi_{\rm eff}$) and precessing ($\chi_{\rm p}$) spins, initial angular momentum $p^0_\phi$, and initial energy $E_0/M$.
        While the estimates for all common parameters differ to varying degrees, the largest discrepancy is in $\chi_{\rm eff}$, with a relative shift of $\epsilon_{\chi_{\rm eff}}=1.416$; the precessing analysis favors lower values, including 0 in the 90$\%$ \ac{ci} and with a longer tail toward the negative axis.
        For GW190521, the hyperbolic scenario is favored over the precessing one with $\ln \mathcal{B}^{\rm hyp}_{\rm prec} = 3.71^{+0.11}_{-0.11}$.
        }
\end{figure*}

Figure~\ref{fig:gw190521} shows the results of our two main analyses for GW190521.
The teal distributions represent the posterior obtained when assuming hyperbolic, non-precessing systems, and the orange distributions represent the posterior when assuming \ac{QC}, precessing systems.
The top panels of each column show the 1D marginalized distribution for the corresponding parameter, with the $90\%$ \ac{ci} marked by the dashed lines.
The other panels show the 2D distributions for the corresponding pair of parameters, with the contours bounding the $50\%$ and $90\%$ credible regions.
When comparing the two analyses, we find a significant shift in multiple parameters.
We first find a noticeable difference in the total (source frame) mass and, to a lesser extent, the mass ratio.
The relative shifts for the total mass and mass ratio are $\epsilon_{M_{\rm source}}=0.809$ and $\epsilon_q=0.220$, respectively.
Concerning the spins, we find a large difference in the results for the effective aligned spin $\chi_{\rm eff}$, with a relative shift $\epsilon_{\chi_{\rm eff}}=1.416$.
This substantial difference is at least partly due to the \ac{QC} analysis returning some support for a spin-precessing description, with larger perpendicular components favored for the spin vectors; the median effective precessing spin parameter we find is $\chi_{\rm p} = 0.54^{+0.33}_{-0.35}$.
As for the hyperbolic-specific parameters $E_0, p_\phi^0$, we find that both rail against their priors' bounds.
We recover more support for lower energies, with a median value $E_0 = 1.06_{-0.05}^{+0.09}$, and for lower angular momenta, with median $p_\phi^0 = 2.69_{-0.62}^{+1.61}$.
The posterior-supported region of the $E_0-p_\phi^0$ parameter space thus corresponds to direct plunge waveforms.

As seen in Table~\ref{tab:events}, GW190521 is the only event from our analyses that favors the hyperbolic scenario over the \ac{QC}, precessing one, with \ac{BF} $\ln \mathcal{B}^{\rm hyp}_{\rm prec} \simeq 3.71^{+0.11}_{-0.11}$.
A plethora of claims have been put forward in the literature about the origin of GW190521~\cite{LIGOScientific:2020iuh,2020ApJ...900L..13A,Gamba:2021gap,2020ApJ...903L...5R,PhysRevD.109.024024,Iglesias:2022xfc,Bustillo:2021tga,PhysRevLett.126.081101,Lai:2025skp,jxrc-z298,Gayathri:2020mra,Gayathri:2020coq}.
To strengthen our result and investigate in more depth the extent to which the hyperbolic description is favored over alternative scenarios, we take advantage of the flexibility of \dali~to analyze this event under multiple additional hypotheses: \ac{QC}, non-precessing; eccentric, non-precessing; and eccentric, precessing.
We also compare our \dali~runs with an analysis performed with the \ac{QC}, precessing NR surrogate model \nrsur~\cite{2019PhRvR...1c3015V}, since the main result from the original \ac{LVK} analysis of GW190521 was obtained with it.
\acp{BF} comparing these analyses are presented in Table~\ref{tab:small}.
Besides being favored over the \ac{QC}, precessing scenario, the hyperbolic scenario is also preferred over all other configurations, including that of a system featuring both non-zero eccentricity and precession.
Indeed, all of the alternative scenarios we considered are \emph{more} disfavored compared with the hyperbolic one than our original \ac{QC}, precessing analysis.
This likely reflects a preference for spin-precessing systems and little support for non-circular dynamics in the bound case; eccentric analyses are thus penalized in \ac{BF} computations because of their larger prior volume.
Finally, we find a slight preference for \dali's hyperbolic description of the event over \nrsur's \ac{QC}, precessing one, although not as significant, with $\ln \mathcal{B}^{\rm hyp}_{\rm NRSur} \simeq 2.01^{+0.11}_{-0.11}$.

\subsection{GW231123}
\label{subsec:gw231123}
Another short, high-mass event -- with the highest total mass inferred to date for a \ac{GW} signal -- GW231123 has been found to match best to highly precessing, \ac{BBH} systems on \ac{QC} inspirals, typically with analyses showing support for nearly extremal spin magnitudes~\cite{LIGOScientific:2025rsn}.
The event is also notable as one whose source properties are most affected by systematic uncertainty, due to significant differences between estimates obtained using different waveform models~\cite{gxjb-23lv,ly9b-w75v}.

\begin{figure*}
    \includegraphics[width=0.85\textwidth]{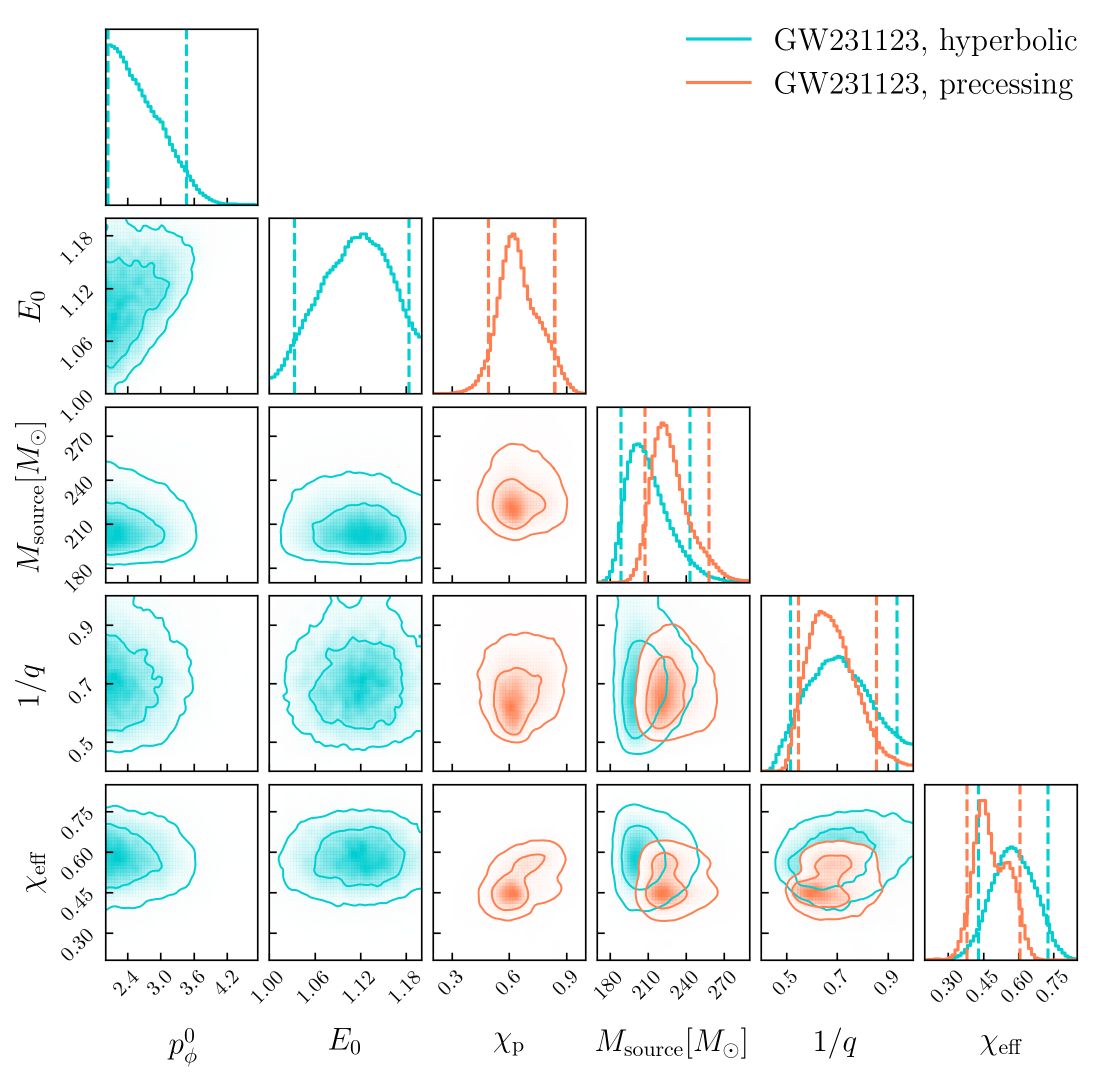}
    
    \caption{
        \label{fig:gw231123}
        \textbf{Hyperbolic and precessing analyses for GW231123:} This figure compares the results of the hyperbolic (teal) and precessing (orange) analyses for GW231123 using~\dali.
        The panels at the top of each column are the 1D marginalized posterior distributions for each parameter, while the others show the 2D posteriors and the contours of the $50\%$ and $90\%$ credible regions.
        The parameters shown are the mass ratio $q$, source-frame total mass $M_{\rm source}$, effective aligned ($\chi_{\rm eff}$) and precessing ($\chi_{\rm p}$) spins, initial angular momentum $p^0_\phi$, and initial energy $E_0/M$.
        As for GW190521, the most significant difference we find between the two analyses is in the inferred $\chi_{\rm eff}$ ($\epsilon_{\chi_{\rm eff}}=0.797$), for which the precessing \ac{PE} favors lower values (although still excluding 0 in this case).
        Notably, for GW231123, the estimated total mass is higher in the precessing case, contrary to what we obtained for GW190521.
        The hyperbolic scenario is not favored over the precessing scenario in this case, with $\ln \mathcal{B}^{\rm hyp}_{\rm prec}=-15.80^{+0.24}_{-0.24}$ for GW231123.
        }
\end{figure*}

As a short, ringdown-dominated, high-mass signal somewhat similar to GW190521, GW231123 seemed to be another good candidate for a description as a hyperbolic encounter resulting in merger.
As shown in Table~\ref{tab:small}, we find that this is not the case.
With a $\ln \mathcal{B}^{\rm hyp}_{\rm prec}=-15.80^{+0.24}_{-0.24}$, this event strongly favors the \ac{QC}, precessing scenario.
Figure~\ref{fig:gw231123} shows the posterior distributions of the hyperbolic analysis in teal and the \ac{QC}, precessing one in orange. Relative to the \ac{QC}, precessing analysis, the inclusion of hyperbolic parameters with non-precessing spins shifts the posterior to lower values of $M_{\rm source}$, higher values of $\chi_{\rm eff}$, and a more broader posterior for $q$. Similar to our result for GW190521, we find a railing posterior at the lower bound of $p^0_\phi$. Unlike GW190521, we find the recovered $E_0$ posterior to favor higher energies peaking around 1.12. As described in further detail in Section~\ref{subsec:hyp-parms}, we suspect this support for higher energies is due to the fact that these systems lack any orbits and just result in a direct plunge (i.e. only have a ringdown). Since the current version of the hyperbolic configuration of~\dali~includes a \ac{QC} ringdown prescription, we speculate that this allows the PE to match better to higher energy systems. Conversely, lower energy systems, which include some orbits before plunging, do not match well to GW231123 (in contrast to GW190521; see Section~\ref{subsec:gw190521}).

Like for GW190521, for this event we again perform additional \ac{PE} runs with~\dali~incorporating orbital eccentricity and/or spin-precession, as well as with~\nrsur~in the \ac{QC}, precessing scenario.
\acp{BF} comparing the various hypotheses with the hyperbolic analysis are collected in Table~\ref{tab:small}.
For GW231123, we find that the inclusion of spin-precession in the analysis is key for high evidences.
The descriptions most favored by the data are the \ac{QC}, precessing and, to a lesser extent, the eccentric, precessing scenario. While not a focus here, we find consistent results with previous eccentric analyses on GW231123~\cite{ly9b-w75v}. Out of the non-precessing cases (hyperbolic, eccentric, and \ac{QC}), the hyperbolic scenario is the preferred one.
An analysis combining spin-precession with hyperbolic dynamics would be interesting, but this is not supported by~\dali.
Its prescription to adapt the twist procedure~\cite{Schmidt:2010it,OShaughnessy:2011pmr,Schmidt:2012rh,Pekowsky:2013ska,Akcay:2020qrj,Gamba:2021ydi} for precessing waveforms to the bound, non-circular case~\cite{Gamba:2024cvy} is unlikely to accurately model hyperbolic trajectories, as it relies on orbit-averaged \ac{pn} equations of motion for the angular momentum vectors, also notably not including explicit non-circular corrections.


\subsection{Follow-up maximum likelihood synthetic injections}
\label{subsec:synthetic}
To investigate our ability to actually measure evidence of hyperbolic systems, we carry out analyses of two synthetic signals for each of GW190521 and GW231123: the maximum likelihood waveform from both the hyperbolic and the precessing analyses.
Table~\ref{tab:synthetic} lists the parameters of the four waveforms.
We project the GW190521-like signals into a three detector network comprising LIGO-Hanford, LIGO-Livingston, and Virgo; we consider only the LIGO detectors for the GW231123-like signals, to match the real event.
We add no noise when projecting each signal onto the detectors, so as to focus on studying actual biases and systematics, without random noise fluctuations.
For all these injections, we perform PE with~\dali~in the hyperbolic and \ac{QC}, precessing configurations.

\begin{figure*}
    \includegraphics[width=0.495\textwidth]{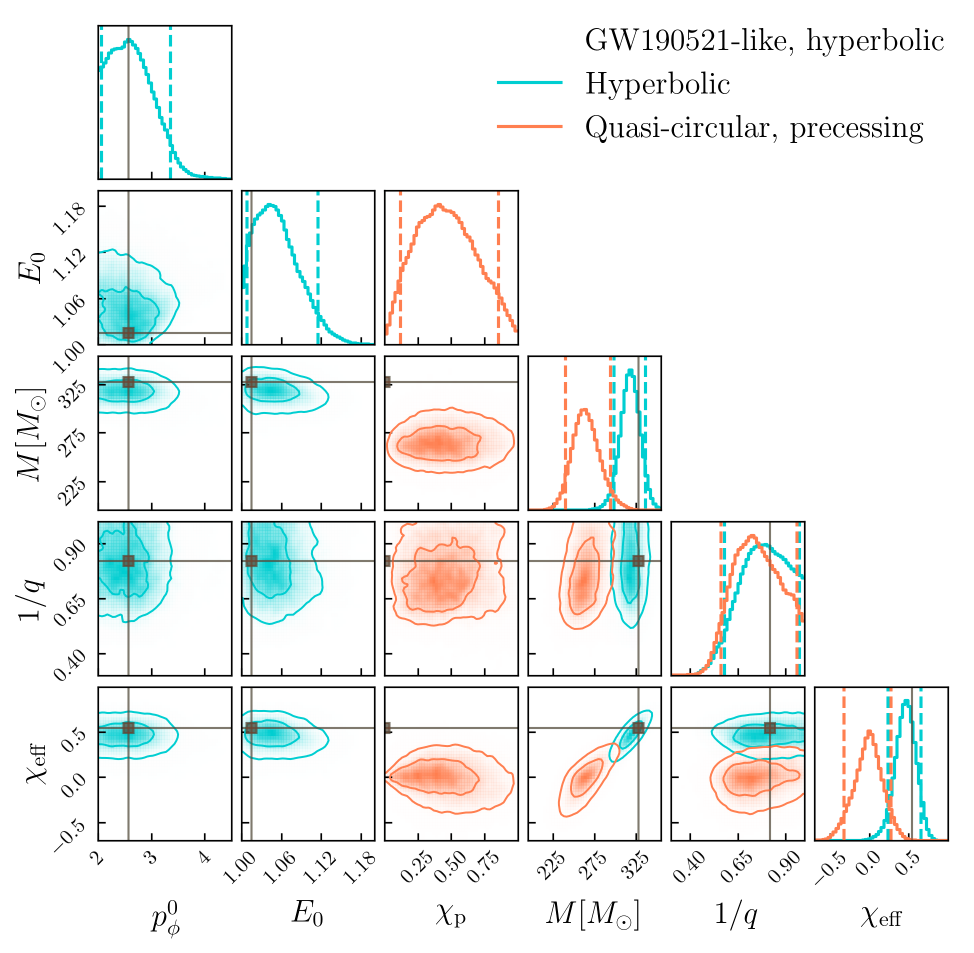}
    \includegraphics[width=0.495\textwidth]{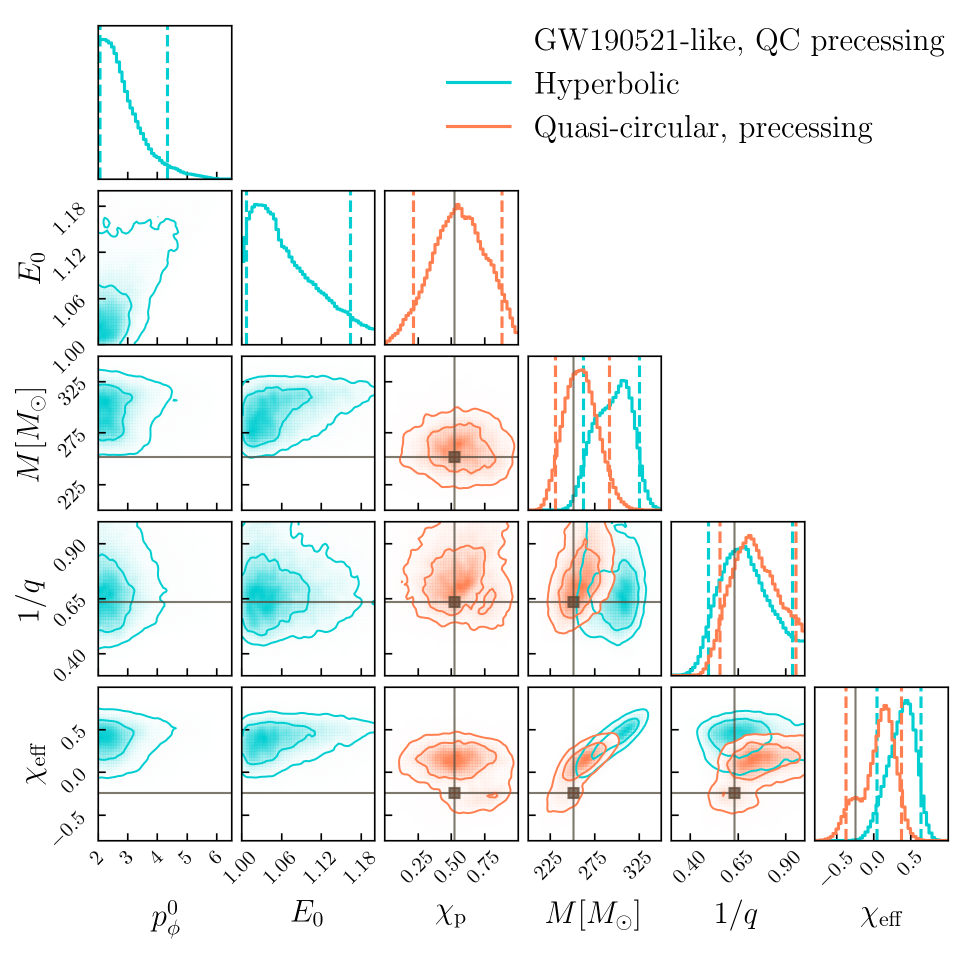}
    
    \caption{
        \label{fig:gw190521-like}
        \textbf{GW190521-like injections:} This figure displays the results of the hyperbolic (teal) and precessing (orange) analyses of the synthetic signals generated from the maximum likelihood samples found in the hyperbolic (left) and precessing (right) \ac{PE} of GW190521.
        We use the \dali~model for both generating the injections and recovery.
        For the detector-frame total mass $M$, (inverted) mass ratio $1/q$, effective aligned and precessing spin parameters $\chi_{\rm eff}, \chi_{\rm p}$, initial energy $E_0$ and angular momentum $p_\phi^0$, we show the 1D marginalized posterior distributions in the top panel of each column.
        The rest of the plots represent the 2D distributions with their 50\% and 90\% credible regions.
        Dark grey lines and markers show the injected true values of each parameter.
        }
\end{figure*}

\begin{figure*}
    \includegraphics[width=0.495\textwidth]{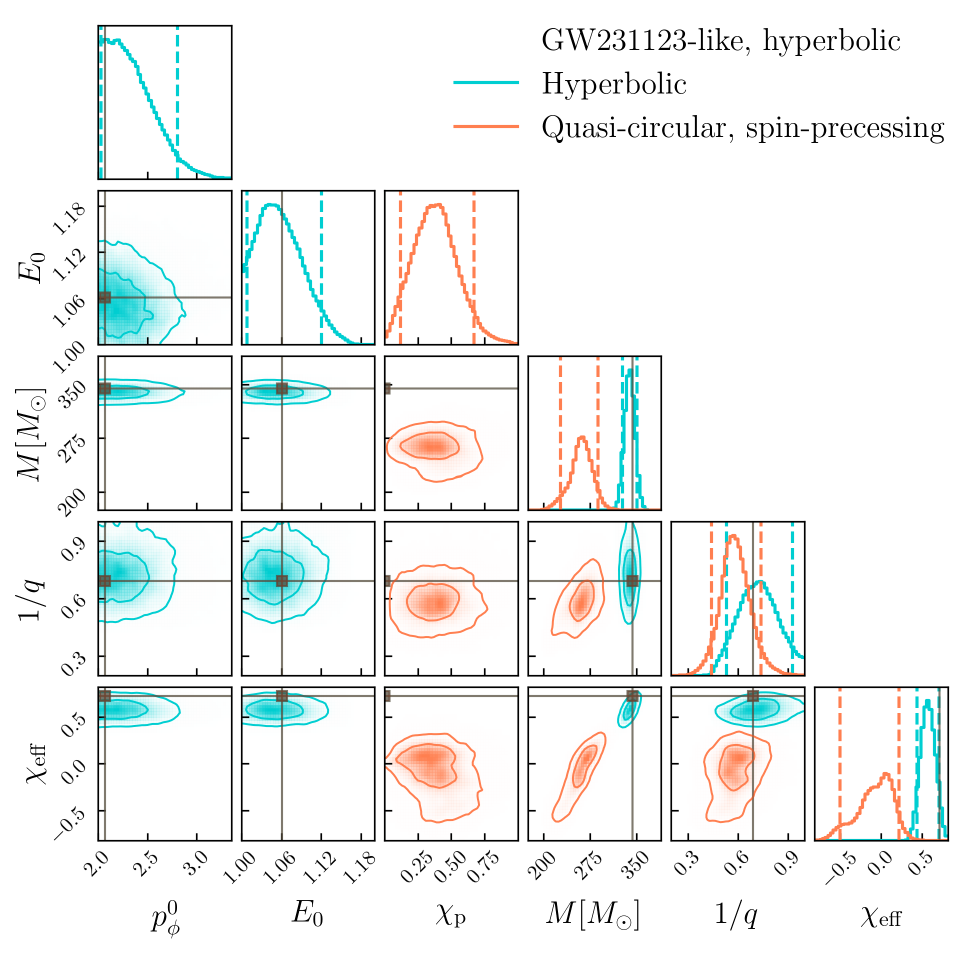}
    \includegraphics[width=0.495\textwidth]{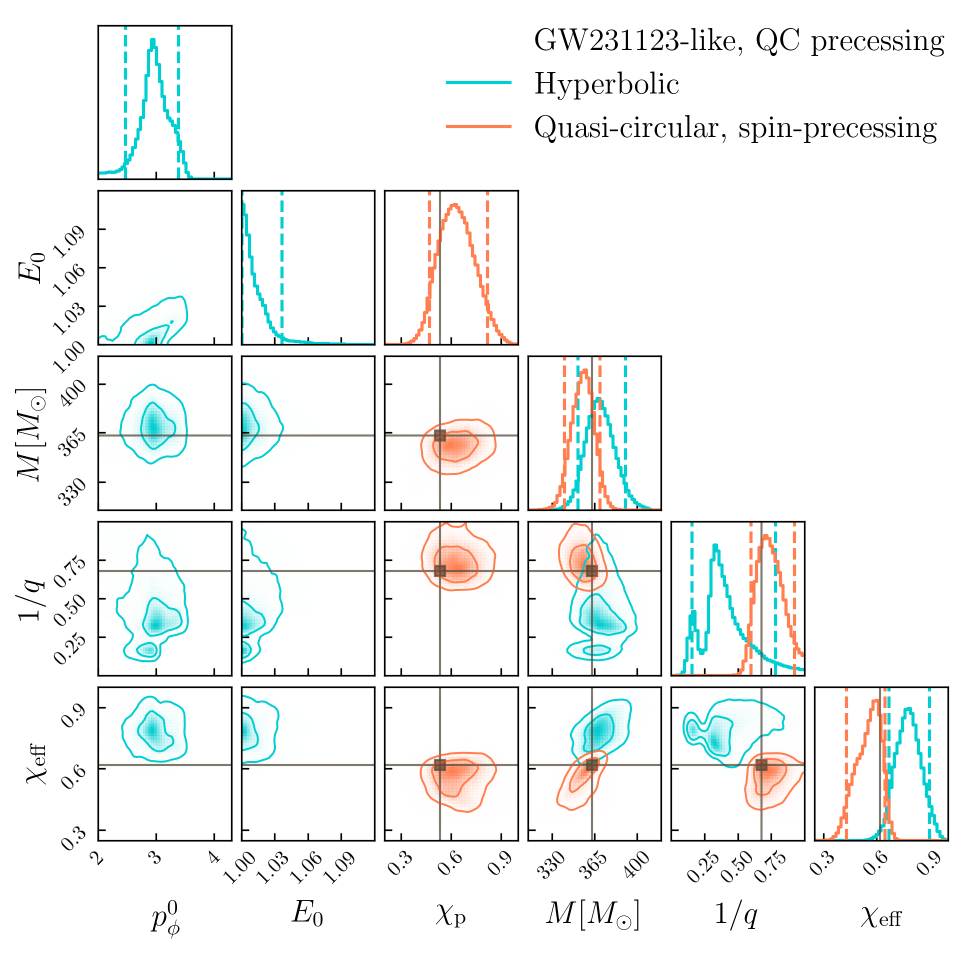}
    
    \caption{
        \label{fig:gw231123-like}
        \textbf{GW231123-like injections:} This figure compares the posteriors from the hyperbolic (teal) and precessing (orange) analyses of the hyperbolic (left) and precessing (right) GW231123-like injections, each based on the maximum likelihood sample obtained in the corresponding real analysis.
        We generate the injected waveforms and analyze them with the \dali~model.
        The panels at the top of each column show the 1D marginalized posterior distributions for each parameter; the other panels represent the $50\%$ and $90\%$ credible regions in the 2D distributions.
        The parameters shown are the detector frame total mass $M$, (inverted) mass ratio $1/q$, effective spins $\chi_{\rm eff}$ and $\chi_{\rm p}$, initial angular momentum $p^0_\phi$, and initial energy $E_0/M$.
        The dark gray lines and markers represent the injected true values for each parameter.
        }
\end{figure*}

The results of these analyses are presented in Figures~\ref{fig:gw190521-like} and~\ref{fig:gw231123-like}, for the GW190521-like and GW231123-like signals, respectively.
For all four of the injections, the matched analyses (meaning, those carried out with the same \ac{BBH} configuration as the injected signal) correctly recovered the true values of most of the parameters, with only the $\chieff$ posterior of the matched hyperbolic GW231123-like run lying just outside the 90\% \ac{ci}.
Due to the extremal spins of the injection and moderate prior assumptions used in the analysis, the shift of the posterior away from the true value is unsurprising.
As expected, the mismatched analyses' results (i.e., hyperbolic analyses of a precessing signal and vice versa) are instead biased from the true values, in some cases severely.
For $\chi_{\rm eff}$, the mass ratio, and the total mass, \ac{QC}, precessing analyses of hyperbolic signals consistently return posteriors shifted toward lower values than the true.
Mirroring this, results from hyperbolic analyses of \ac{QC}, precessing waveforms are biased to higher values for the cases of $\chieff$ and $M$.
Estimates of $q$ in these analyses do not follow a similarly consistent pattern; the multimodal, highly biased posterior obtained in the mismatched analysis of the precessing GW231123-like injection particularly stands out.

To quantify our ability to measure evidence in support of the correct scenario, we calculate the \acp{BF} contrasting each pair of analyses; these are listed in Table~\ref{tab:small}.
For the GW231123-like signals, we find in each case strong preference for the correct scenarios, suggesting these waveforms are significantly different from each other.
As mentioned in Section~\ref{subsec:gw231123}, a strong preference for precessing systems seems to emerge from the data for GW231123.
We find only marginal preference instead for the matched analyses of the GW190521-like injections, meaning that these signals are morphologically similar to each other, and harder to tell apart in \ac{PE}.
These results suggest that there exist regions of the hyperbolic parameter space where waveforms can to some extent resemble signals from certain \ac{QC}, precessing systems, and that GW190521 might belong to such regions.
While the weak preference for the correct scenarios for the GW190521-like injections might diminish one's confidence in our hyperbolic interpretation of the event, we emphasize that this result is anecdotal.
The posterior, see Figure~\ref{fig:gw190521}, in the $p_\phi^0,E_0$ plane is relatively broad, so it's possible other high likelihood points would yield a higher \ac{BF} in favor of the hyperbolic scenario (the GW190521 \ac{BF} was itself higher than the fake injection, see Table~\ref{tab:small}).
A more detailed injection study would be needed to make any systematic claims about degeneracies between hyperbolic and \ac{QC} signals. Another potential direction for future work would be to investigate if, and when, these findings remain valid in the context of third generation detectors and their improved sensitivity at lower frequencies.

\begin{figure}
    \includegraphics[width=\columnwidth]{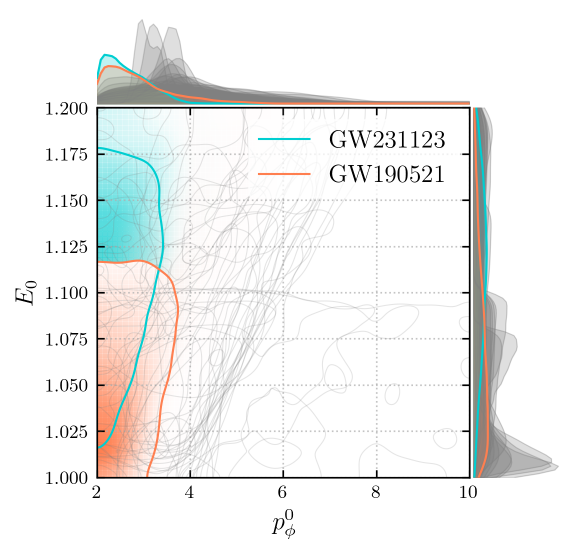}
    
    \caption{
        \label{fig:hyp-params}
        \textbf{Initial energy and angular momentum for all GW events:} This figure compares the $E_0, p_\phi^0$ posteriors recovered when using hyperbolic waveforms for all \totalevents events analyzed.
        We highlight the 2D distributions for GW190521 (teal) and GW231123 (orange), plotting all other events in gray; contours mark the 90\% credible regions.
        The top and side panels show the 1D marginalized posterior distributions, with 90\% \acp{ci} bounded by dashed lines for the highlighted events.
        In most cases, distributions are broad and find no support for a hyperbolic encounter.
        For GW190521, we infer support for both low initial energy and initial angular momentum; this is the only signal that favors a hyperbolic description.
        GW231123 returns even higher preference for low $p_\phi^0$, but a much broader $E_0$ distribution.
        }
\end{figure}

\begin{table*}
\begin{tabular}{l l c c c c c c c c c c c c c c}
    \toprule
    \textbf{Event} &  & \bmath{$M [M_{\odot}]$} & \bmath{$1/q$} & \bmath{$\chi_{\mathrm{1x}}$} & \bmath{$\chi_{\mathrm{1y}}$} & \bmath{$\chi_{\mathrm{1z}}$} & \bmath{$\chi_{\mathrm{2x}}$} & \bmath{$\chi_{\mathrm{2y}}$} & \bmath{$\chi_{\mathrm{2z}}$} & \bmath{$E_0$} & \bmath{$p_{\varphi,0}$} & \bmath{$\iota$} & \bmath{$\psi$} & \bmath{$d_L$} & \bmath{$\phi_{\mathrm{ref}}$} \\
    \midrule \\[-0.7em]
\multirow{2}{2.5cm}[-0.5ex]{GW190521          } & Precessing & $251.42$ & $0.63$ & $0.85$ & $0.24$ & $-0.47$ & $-0.02$ & $-0.09$ & $0.11$ & & & $-0.990$ & $-0.22$ & $1989.30$ & $3.55$ \\[0.3em]
     & Hyperbolic & $327.60$ & $0.82$ & & & $0.96$ & & & $0.04$ & $1.015$ & $2.57$ & $0.784$ & $2.55$ & $3971.73$ & $4.06$ \\[0.3em]
     \multirow{2}{2.5cm}[-0.5ex]{GW231123          } & Precessing & $363.03$ & $0.68$ & $-0.74$ & $-0.38$ & $0.42$ & $-0.11$ & $-0.04$ & $0.91$ & & & $0.771$ & $2.77$ & $4030.96$ & $3.63$ \\[0.3em]
     & Hyperbolic & $344.11$ & $0.69$ & & & $0.59$ & $$ & $$ & $0.91$ & $1.061$ & $2.07$ & $0.759$ & $2.91$ & $3170.69$ & $4.01$ \\[0.3em]
\bottomrule
\end{tabular}
\caption{\label{tab:synthetic}
\textbf{Maximum likelihood injection parameters:} Maximum likelihood points from the hyperbolic and \ac{QC}, precessing analyses of GW190521 and GW231123, which we use to generate synthetic waveforms for our injection-recovery study.
For all injections, the coalescence time is taken to be the corresponding maximum likelihood time for each analysis.
The columns report the event name, configuration, detector frame total mass $M$, inverted mass ratio $1/q$, spin components $\chi_{k, a}$, initial energy $E_0$ and angular momentum $p_\phi^0$, inclination $\iota$, polarization angle $\psi$, luminosity distance $d_L$, and reference phase $\phi_{\rm ref}$.
}
\end{table*}

\section{Conclusions}
\label{sec:conclusions}
We presented a re-analysis of \totalevents high-mass \ac{GW} events that potentially feature \acp{BH} in the \textit{pair-instability mass gap} using a \ac{GW} model for binaries on initially unbound orbits:~\dali.
To quantify the significance of this analysis, we also ran with the bound-orbit, \ac{QC}, precessing version of \dali~and calculated \acp{BF} between the two hypotheses.
All of the events, including GW231123, favored the \ac{QC}, precessing scenario to varying degrees, except for GW190521.
This event favored a hyperbolic morphology with $\ln \mathcal{B}^{\rm hyp}_{\rm prec}=3.71^{+0.11}_{-0.11}$.
Although GW231123 is in a number of ways reminiscent of GW190521, its \ac{PE} results are dramatically different, with it strongly favoring the \ac{QC}, precessing description over the hyperbolic one: $\ln \mathcal{B}^{\rm hyp}_{\rm prec}=-15.80^{+0.24}_{-0.24}$.

As seen in Figure~\ref{fig:violin}, we observed noticeable shifts in the posteriors between the hyperbolic and precessing analyses.
No systematic differences are clearly visible, aside from a slight tendency by the precessing analyses to favor lower total (source-frame) masses and effective aligned spins; the latter effect can at least in part be attributed to known behavior in the comparison of precessing and non-precessing analyses.
We found the largest relative shift for the source-frame total mass and mass ratio in the results for GW230914\_111401, with $\epsilon_{M_{\rm source}}=1.57$, and GW231028\_153006, with $\epsilon_q=0.708$, respectively.
The largest relative shift for $\chi_{\rm eff}$ we obtained instead from GW1901109\_010717, with $\epsilon_{\chi_{\rm eff}}=4.142$.
In the hyperbolic analyses, we found uninformative posteriors for many of the events that span the full prior range of $E_0$, as seen in Figure~\ref{fig:hyp-params}.
As explained in Section~\ref{subsec:settings}, this prior range was chosen mostly due to the limitations of the model.
There are also subsets of events that either favored lower or higher energies.
\ac{PE} results in the latter category should be interpreted with care (see last paragraph of Section~\ref{subsec:hyp-parms} for more details).

We further built confidence in our results for GW190521 and GW231123 by carrying out additional analyses under different hypotheses: \ac{QC}, non-precessing; eccentric, non-precessing; eccentric, precessing, all using \dali; and \ac{QC}, precessing with the \nrsur~model.
The hyperbolic hypothesis remained the one favored by the data in the case of GW190521 across all of these, including the \ac{NR} surrogate run (although with smaller \ac{BF} compared to~\dali~\ac{PE}).
Modeling spin-precession was instead found to be the key factor to obtain high evidences when analyzing GW231123: the \acp{BF} comparing the hyperbolic analysis with all scenarios including precession had the latter favored, with $\ln \mathcal{B}^{\rm hyp}_{\rm prec}=-15.80^{+0.24}_{-0.24}$ and $\ln \mathcal{B}^{\rm hyp}_{\rm e+p}=-12.24^{+0.26}_{-0.26}$.
In contrast, the hyperbolic scenario is significantly favored over the other non-precessing configurations.
Spin-precession on unbound orbits is not supported by \dali~at the time of writing.
Our results suggest that such a configuration could fit the data better, although the phenomenology has seldom been studied with significant development work needed to understand and model the spin dynamics of a direct plunge.

Finally, we analyzed simulated signals generated from the maximum likelihood samples from both the hyperbolic and \ac{QC}, precessing analyses of GW190521 and GW231123.
The injected \ac{BBH} parameters were accurately recovered in matched analyses, while we obtained severely biased results with unmatched signal/waveform configurations.
The GW231123-like injections strongly favored the correct description (e.g., the \ac{BF} for the hyperbolic signal strongly favored hyperbolic systems); conversely, we found that the GW190521-like injections did so only marginally.
This could imply that, while signals from unbound and bound, precessing \ac{BBH} systems are typically clearly distinguishable in \ac{PE} -- especially at lower masses, but also in the parameter space neighborhood of GW231123 -- the GW190521 event might represent an example of a binary configuration where high total mass and average \ac{snr} make the two scenarios harder to tell apart.
Since the $p_\phi^0, E_0$ posterior is relatively broad, a more thorough investigation of the parameter space would be needed to establish whether these results are anecdotal or indicative of such a degeneracy for systems like GW190521.
Moreover, we note that the \ac{BF} we find in the analyses of the real event is noticeably larger than that comparing the synthetic injections'.
In a forthcoming follow-up paper, we intend to carry out a systematic injection-recovery campaign to fully understand our ability to measure \ac{BBH} parameters from hyperbolic signals across the parameter space.

While we did not find a population of hyperbolic encounters in the pair-instability mass gap, we did obtain noticeable evidence in favor of the hyperbolic scenario for GW190521.
Nominally, our result is not as significant as that reported in~\citet{Gamba:2021gap}, although the many differences in our setups need to be taken into account when comparing the two studies.
Compared to~\citet{Gamba:2021gap}, our analyses newly include: (i) higher order modes, (ii) non-precessing spins, (iii) a larger prior range for the initial energy and angular momentum, and (iv) an improved version of the~\dali~model.
The wider explored parameter space should naturally lead to lesser statistical evidence in favor of the hyperbolic hypothesis; the fact that a new analysis performed with a more complete model still returns significant preference for it corroborates this interpretation.
Our additional analyses in all configurations supported by~\dali~provide further evidence in this direction.
An important caveat to this discussion is the lackluster faithfulness of \dali~in the high-energy regime (not excluded by our GW190521 posterior estimate), in both its underlying dynamics and waveform model.
In particular, this study should be updated once \ac{NR} information from non-circular systems is incorporated into its merger-ringdown prescription, and a more robust treatment of the transition from plunge to merger is developed.
In summary, our result, in tandem with previous~\cite{Gamba:2021gap} and contemporary~\cite{Pompili:v6pe} works, builds the case for GW190521 being an initially unbound system; however, improvements to the \dali~model and a better understanding of hyperbolic \ac{PE} are needed to further strengthen our confidence in the interpretation of our result.

In parallel to the present work, Pompili \textit{et al.} in~\cite{Pompili:v6pe} carried out \ac{PE} under the unbound orbit hypothesis for a small sample of events using the \texttt{SEOBNRv6EHM} model, including the GW190521, GW190620, GW191109\_010717 and GW231123 events which we also consider here.
This work qualitatively agrees with our results for GW190521 and GW231123; however, our findings are significantly different for GW191109\_010717, which they find is best described as a direct plunge.
Further studies are clearly needed to fully develop waveform models and \ac{PE} techniques to adequately treat the case of unbound orbits.
In future work, we will present an injection-recovery study across the hyperbolic parameter space, to more deeply investigate our ability to measure the properties of these systems.
We also plan a follow-up exploration of the extent of the systematic uncertainties in \ac{PE} due to differences between hyperbolic waveform models, also making contact with the available \ac{NR} simulations in this regime.
These additional studies will be vital not only to fully assess the potential intrinsic limitations of \ac{PE} involving systems on unbound orbits, but also to build confidence in the results of this work.

\acknowledgements
The authors would like to thank Rossella Gamba, Daniel Williams, Sebastiano Bernuzzi, and Jacopo Tissino for their suggestions and discussions.
JL and DC acknowledge support from the Italian Ministry of University and Research (MUR) via the PRIN 2022ZHYFA2, GRavitational wavEform models for coalescing compAct binaries with eccenTricity (GREAT). 
ROS acknowledges support from NSF awards PHY 2012057 and 2309172.
LC and CH acknowledge support from NSF awards PHY 2110481 and PHY 2409714.
PL acknowledges support from Phenikaa University grant No. PU2024-3-A-05.
The authors are grateful for computational resources provided by the IGWN-OSG and LIGO Laboratory and supported by National Science Foundation Grants PHY-0757058 and PHY-0823459. 
This research has made use of data or software obtained from the Gravitational Wave Open Science Center (gwosc.org), a service of the LIGO Scientific Collaboration, the Virgo Collaboration, and KAGRA.
This material is based upon work supported by NSF's LIGO Laboratory which is a major facility fully funded by the National Science Foundation, as well as the Science and Technology Facilities Council (STFC) of the United Kingdom, the Max-Planck-Society (MPS), and the State of Niedersachsen/Germany for support of the construction of Advanced LIGO and construction and operation of the GEO600 detector.
Additional support for Advanced LIGO was provided by the Australian Research Council.
Virgo is funded, through the European Gravitational Observatory (EGO), by the French Centre National de Recherche Scientifique (CNRS), the Italian Istituto Nazionale di Fisica Nucleare (INFN) and the Dutch Nikhef, with contributions by institutions from Belgium, Germany, Greece, Hungary, Ireland, Japan, Monaco, Poland, Portugal, Spain.
KAGRA is supported by Ministry of Education, Culture, Sports, Science and Technology (MEXT), Japan Society for the Promotion of Science (JSPS) in Japan; National Research Foundation (NRF) and Ministry of Science and ICT (MSIT) in Korea; Academia Sinica (AS) and National Science and Technology Council (NSTC) in Taiwan.
\appendix

\end{document}